# Citation accuracy, citation noise, and citation bias:

# A foundation of citation analysis

Lutz Bornmann* & Christian Leibel*,**

* Science Policy and Strategy Department

Administrative Headquarters of the Max Planck Society

Hofgartenstr. 8,

80539 Munich, Germany.

Email: bornmann@gv.mpg.de, christian.leibel.extern@gv.mpg.de

ORCiD: https://orcid.org/0000-0003-0810-7091

** Department of Sociology

LMU Munich

Konradstr. 6,

80801 Munich, Germany.

ORCiD: https://orcid.org/0009-0006-6893-4901

**Abstract**

Citation analysis is widely used in research evaluation to assess the impact of scientific papers. These analyses rest on the assumption that citation decisions by authors are accurate, representing the flow of knowledge from cited to citing papers. However, in practice, researchers often cite for reasons that are not related to the fact that there has been (intellectual) input from previous papers. Citations made for rhetorical reasons or without reading the cited work compromise the value of citations as instrument for research evaluation. Past research on threats to the accuracy of citations has mainly focused on citation bias as the primary concern. In this paper, we argue that citation noise – the undesirable variance in citation decisions – represents an equally critical but underexplored challenge in citation analysis. We define and differentiate two types of citation noise: citation level noise and citation pattern noise. Each type of noise is described in terms of how it arises and the specific ways it can undermine the validity of citation-based research assessments. By conceptually differing citation noise from citation accuracy and citation bias, we propose a framework for the foundation of citation analysis. We discuss strategies and interventions to minimize citation noise, aiming to improve the reliability and validity of citation analysis in research evaluation. We recommend that the current professional reform movement in research evaluation such as the Coalition for Advancing Research Assessment (CoARA) pick up these strategies and interventions as an additional building block for careful, responsible use of bibliometric indicators in research evaluation.

# 1 Introduction

Scientific knowledge in all disciplines is a collective endeavor, shaped by researchers who identify research problems and develop solutions within their communities (Aman & Gläser, 2025). Continuous knowledge flow is essential. Researchers usually publish their results in papers that are made available to other researchers. In addition to the author's research results, each paper includes citations. The term 'citation' denotes two related yet conceptually distinct phenomena that are bibliographic references listed at the end of the paper and in-text citation statements within the narrative of the paper (Enkhbayar, 2025). Every paper is linked to previous research via the author's citation links. The collective of citing and cited papers forms a social citation system of formal scientific communication. In the Social Systems Citation Theory (SSCT) proposed by Tahamtan and Bornmann (2022), citation links weave a web of knowledge connecting research papers that represent the entire science system.

In the formal communication process, citations in papers are intended to clarify which ideas, methods, results etc. do not originate from the citing authors themselves but have already been dealt with in previous publications. The process of citation in the social citation system is of particular importance in the evaluation of research (in the attribution of reputation in science): "Citations are a form of scientific currency, actively conferring or denying value" (Penders, 2018). Nowadays, there is hardly any institutional evaluation procedure in science that does not rely on citation analysis. Evaluative citation analysis rests on the assumption that citations can be interpreted, at least approximately, as signals of knowledge flow from cited to citing papers. In this framework, knowledge flow occurs when authors draw on prior work in producing a new publication. When such citation decisions are aggregated at the level of individual papers, publications can be compared according to their intellectual influence on subsequent research. Because paper-level citation data can be further aggregated, citation-based indicators are also used to assess larger units, including researchers, journals, institutions, and countries. The validity of such assessments depends on the extent to which citations consistently function as interpretable signals of knowledge flow between cited and citing publications.

The citation process can be examined from multiple perspectives, as we discuss below. The perspective adopted here is that of evaluative citation analysis in bibliometrics. From this perspective, the central question is: under what conditions can citations, understood as the outcomes of citation decisions, be interpreted as reliably indicating the significance or influence of research? This question has become increasingly important because citations play a central role in research evaluation. As Lawson et al. (2025) note: "With the rise of citation-count indices in evaluating the performance of scientists, citations themselves are, more than ever before, a currency of the scientific realm … Citations are now not merely an index; they are a target". Against this backdrop, we ask what assumptions must hold if citation data are to be used validly in research evaluation. From the perspective of evaluative citation analysis, citation decisions need to be sufficiently accurate in the sense that citations correspond to knowledge flow from cited to citing work (Aksnes et al., 2019). We conceptualize knowledge flow as the process by which an author incorporates insights from another publication – such as methodological contributions or substantive findings – into their own work.

The idea that citations represent knowledge flow and can be accurate or inaccurate in this sense is central to citation analysis, but it often remains implicit. One goal of this paper – in addition to discussing citation noise and bias – is to make explicit that citation analysis implicitly assumes a normative theory of citation. This theory is a set of expectations about how citations should function if they are to serve as valid indicators of intellectual influence. Foundational works in the sociology of science by Robert K. Merton are core to this normative perspective. According to Merton (1973), citation decisions should reflect the communal and cumulative nature of scientific knowledge by acknowledging the intellectual contributions on which new work is built. Small (2004) draws on Merton (1973) to suggest that a norm of accurate acknowledgment of prior work may be expected in scholarly communication. Penders (2018) similarly notes that well-referenced manuscripts place an argument in its proper knowledge context and thereby support its novelty, value, and visibility.

Lawson et al. (2025) formulate a similar requirement from the perspective of citation-based evaluation: "each citation should represent at least one specific input to the citing author's evidentiary and/or argument chain … Ideally, a reader should be able to rely on the author's references as a representative and unbiased database of the most relevant past literature by which to evaluate the author's contribution" (p. 3). In the framework adopted here, the relevant citation decision is therefore not whether a publication was the first or most important contribution on a given topic, but whether it was actually used by the citing author in developing their own work. A citation is most informative for evaluative citation analysis when it points to the specific publication on which the citing paper directly draws. If factors other than this knowledge flow influenced the citing author's decision (such as the reputation of the journal in which the cited paper was published or the reputation of its author), the citation is not informative for research evaluation. In that case, the citation may signal something real – for example, author reputation or journal visibility – but it does not signal the construct that evaluative citation analysis usually aims to measure: the influence of prior work on subsequent research.

Alongside the normative theory of citation, a research strand mostly in bibliometrics has examined the actual reasons why authors cite, and this literature suggests that citations are often made for reasons that do not align with the normative ideal of documenting knowledge flow. Overviews of the relevant literature by Bornmann and Daniel (2008) and Tahamtan and Bornmann (2018, 2019) show that many reasons for citations could be identified. One of the earliest identified reason to cite is that researchers cite scientific literature primarily to reinforce the credibility and legitimacy of their own claims, aiming to persuade readers of their robustness and validity (Gilbert, 1977). Researchers cite "to defend their claims against attack, advance their interests, convince others, and gain a dominant position in their scientific community" (Bornmann & Daniel, 2008, p. 49). The comprehensive literature on reasons to cite suggests that some disciplines are characterized by the prevalence of a different and broader spectrum of reasons than others. For example, the study by Sula and Miller (2014) reveals that linguistics tends to feature reinforcing citations of prior literature, whereas philosophy typically involves more critical engagement with cited works.

Another research strand (mostly in bibliometrics) dealing with citation decisions has focused on inaccurate citations. Wakeling et al. (2025) published a current overview of studies dealing with inaccurate citations. There is some variance in terminology in these studies: What we refer to as 'inaccurate citation' in this study is also referred to as 'inaccurate quotation' or 'inaccurate reference' in other studies. Inaccurate citations may be quotation errors (the citation "does not support the statement to which it is applied", Wakeling et al., 2025, p. 1397) or reference errors ("the in-text citation ... or the formal reference in the reference list, is incorrect or incomplete", Wakeling et al., 2025, p. 1397). An explanation for why citation errors occur may be the 'lazy author syndrome' (Gavras, 2002), which refers to the practice of citing publications without engaging with the content of the cited work. Chen et al. (2024) denote this syndrome as 'heuristic citation approach', where citations "are made without fully reading or understanding the paper. Heuristic approaches may involve lifting citations directly from the references of another paper, incorporating citations used in one's previous work without knowledge of ongoing developments, or referencing work based only on cursory readings". Both the lazy author syndrome and the heuristic citation approach are thus about cursory or careless citation.

The comprehensive empirical literature on inaccurate citations and the diverse reasons why authors cite highlight the limitations of the normative citation theory as a description of real citation behavior. To address these limitations, several alternative citation theories have been proposed such as the social-constructivist theory of citations (Barnes & Dolby, 1970), citations as markers or symbols (Small, 1978), the reflexive indicator theory (Wouters, 1998, 1999), references as threat signals (Nicolaisen, 2004), and the SSCT (Tahamtan & Bornmann, 2022). A detailed overview of these theories can be found in the most recent attempt for a theory of citations (SSCT) published by Tahamtan and Bornmann (2022). The various theories of citations were developed to explain why researchers cite, what functions citations serve in scholarly communication, and how these underlying motivations shape the validity and interpretation of citation-based indicators.

Although this study aims to establish a foundation for citation analysis, it is not its goal to introduce yet another citation theory. Our contribution highlights that citation-based research evaluation implicitly assumes a world in which citation behaviour of authors aligns with the normative citation theory. It follows logically that the validity and reliability of citation-based research evaluation suffer to the extent that citation behaviour deviates from this assumption. It is a methodological problem that cannot be ignored if one wishes to employ citations in research evaluation. Our goal is therefore to discuss measures that may improve the validity and reliability of citation analysis. Both strands of previous research on citation decisions – reasons to cite and inaccurate citations – point to a phenomenon that we would like to address fundamentally in this paper: the distortion of citation decisions by noise (and bias) as opposed to citation accuracy. Identifying noise and bias in citation decisions can be used to improve citation decisions and thus evaluative bibliometrics.

The use of noise and bias in this study follows definitions of Kahneman et al. (2021). The authors distinguish two forms of error in human judgment: bias, understood as systematic deviation, and noise, understood as undesirable variability. Although Kahneman et al. (2021) mostly illustrate these concepts with examples from the justice system, they emphasize that the same principles apply to other evaluative settings, such as asylum decisions, patent assessments, grant reviews, and grading (Kahneman & Frueh, 2022). In the context of citation decisions (another evaluative setting), past (bibliometric) research has dealt almost exclusively with biased decisions (in the context of the research strand dealing with authors' reasons to cite). Traag and Waltman (2022) define bias "as a direct causal effect that is unjustified". Biased citation decisions occur when factors unrelated to knowledge flow, such as authors' gender, have causal effects on citation outcomes. A current overview of the many factors that can lead to biased citation decisions can be found, for example, in Kousha and Thelwall (2024). We would like to highlight just one example of citation bias here: Jannot et al. (2013) have found that there is "a citation bias favoring [statistically] significant results ... in medical research. As a consequence, treatments may seem more effective to the readers of medical literature than they really are" (p. 296). In this quote, the authors not only name the potential citation bias, but also list the consequences that can arise from this bias: the distorted citation analysis can give a false impression of the state of research.

To understand citation errors in citation decisions, it is necessary to address not only bias but also noise which hardly plays a role in literature. This paper addresses the gap by analyzing noise and bias in citation decisions following the approaches of Kahneman et al. (2021). We show that citation noise constitutes a significant problem in citation decisions and, consequently, in research evaluation, alongside citation bias. However, we were only able to identify a few papers (from bibliometrics) in the Web of Science (WoS, Clarivate) in which the term 'citation noise' appears in the title or abstract (Cawkell, 1969; Meho & Yang, 2007; Tang, 2023; Wei et al., 2019) (date of search: April 2025). The authors of these papers usually use the term 'citation noise' but do not explicitly define it. Even though the term and its associated concept are hardly known in bibliometrics, the term has been in use in the field of patent analysis since the 1990s. In an overview of the literature on this phenomenon, Smith (2014) defines citations as noise in a patent if they do not represent a knowledge flow from the prior art to the patent. For the author, knowledge flow occurs "when a patent is building upon or improving older technology and may be identified when there is direct link between the claims of the patent and relevant citations" (Smith, 2014, p. 40). Results reveal that citations in a patent not representing knowledge flow may account for half of its total citations (Jaffe et al., 2000).

We assume similar numbers for citations in publications: The literature overview by Bornmann and Daniel (2008) of studies on reasons to cite reveals a comparatively frequent occurrence of citations of the perfunctory (up to 50 percent) and persuasive (up to 40 percent) type. Perfunctory citations are defined as superficial, routine or non-essential citations that do not make a substantive contribution to the argument or knowledge base of the citing publication. They are included even though the cited work did not meaningfully influence the research. Persuasive citations are citations that serve a rhetorical or argumentative purpose rather than a substantive epistemic one. They do not advance the conceptual or empirical content of the citing paper but are used to strengthen the author's position, signal alignment with a particular school of thought, or lend authority to an argument. In empirical studies, they are often grouped with perfunctory citations because they do not fulfill a pragmatic knowledge function in the text.

The results of Donner et al. (2025) show that around half of the citations in the WoS are of background type. Background citations (superficially) mention prior research that provides the scholarly context for a study. Teplitskiy et al. (2022) tested whether citations reflect rhetorical usefulness or influence on research. The authors used "data on 17,154 randomly sampled citations collected via surveys from 9,380 corresponding authors in 15 fields" and found that "most citations (54%) had little-to-no influence on the citing authors". Estimates of the prevalence of citation errors vary across studies due to differences in how errors are defined and classified. For example, some studies distinguished between major and minor errors, while others classified citations as fully accurate, partially accurate, and unsubstantiated. These methodological differences may partially explain why reported error rates range from 5% to 40% (Wakeling et al., 2025) with most studies reporting between 10% to 20% (including the study by Wakeling et al., 2025). Wakeling et al. (2025) illustrate these results as follows: "If a typical social science article has, on average, 34 references … the implication is that 3-6 of the citations in the paper will be erroneous in some way" (p. 1397). Two meta-analyses examining inaccurate citations report error rates of 25.4% (Jergas & Baethge, 2015) and 14.5% (Mogull, 2017).

The many results of the studies on perfunctory citations and inaccurate citations reveal that many citations in publications do not represent knowledge flow (but contribute to noise or bias). There seems to be a similar phenomenon in publication citations as in patent citations where citations may not represent knowledge flow in many cases (as reported by Jaffe et al., 2000). Although patent citation noise is seen as "a major challenge to the effectiveness of patent evaluation methodologies, which may therefore be poor indicators of the economic value of patents" (Smith, 2014, p. 40), it is striking that citation noise has so far received comparatively little attention as a challenge in bibliometrics and science policy. However, as we will show in the following, dealing with citation noise is essential for the use of citation analysis in research evaluation. Bibliometric research should not only deal with citation bias but should also address citation noise.

What do we mean by citation noise in bibliometrics, and how can it be distinguished from citation bias and citation accuracy? In the framework adopted here, citation accuracy refers to the extent to which citation decisions reflect knowledge flow from the cited paper to the citing paper. Citation errors can take two forms: noise and bias. Noise refers to the undesirable variability in judgments about which papers should be cited in a particular paper; it represents the (random) scatter of citation decisions among citing authors. In contrast to noise, citation bias is the systematic deviation of citation decisions from accurate citation decisions. For example, if authors preferentially cite papers with statistically significant results, this phenomenon is referred to as citation bias. Studies that do not have statistically significant results tend to be ignored by the citing authors. Both citation noise and citation bias affect aggregated citation indicators, such as citation counts or citation rates, by weakening the extent to which they reflect the intellectual influence of research on subsequent studies.

Citation analyses are used to decide on reputation and resources in nearly all disciplines: various methods are used in research evaluation "with bibliometric data playing an important role in (almost) all evaluations" (Hulek & Teschke, 2023, p. 36). The results of Hammarfelt et al. (2020) on the use of bibliometrics in the peer review process show that "this external and numerical information … plays an important role when candidates are compared and ranked" (p. 43). Since there is a recognized need – both among researchers and within science policy – to use citation analyses for evaluative purposes, citation data should be able to give the right signal for research evaluation. If it is primarily noise and bias that determine citation decisions, then there is a risk that incorrect reputation and resources decisions will be made based on citation data. As we will show in the following, noise and bias can be significant sources of error in citation decisions. We propose a framework for the foundation of citation analysis to make citation analysis function more effectively.

Against the backdrop of citation analysis being potentially distorted by bias and noise, one could take the position that citation-based indicators are unsuitable for research evaluation and should be replaced by peer review or other qualitative assessment procedures. Several reform initiatives discussed in Section 3 raise concerns that point in this direction. While we acknowledge that citation-based evaluation has important limitations, our aim is not to argue for or against the use of citation analysis as such. Instead, we propose a framework for specifying the conditions under which citation analysis can provide valid information for research evaluation. This framework builds on established psychological work on decision-making errors and applies the distinction between accuracy, bias, and noise to citation decisions. Given the continued role of citation data in research evaluation, it is important to examine how accurately and consistently citation decisions reflect the construct that citation analysis is intended to measure. We propose measures that are intended to increase the validity of citation data for citation analysis by eliminating noise and bias in citation decisions.

In the next section (section 2), we introduce our framework with the concepts of citation accuracy, citation noise, and citation bias in detail based on an exemplary social citation system composed of a small fictitious world of cited and citing papers. With the introduction of these basic concepts, we sketch the foundation of a citation analysis that considers citations as knowledge flow proxies. In the following section 3, against the background of noise and bias in citation decisions, we deal with strategies to reduce citation noise and bias.

# 2 Definition and measurement of citation accuracy, citation bias, and citation noise

We would like to explain the definition and measurement of citation accuracy, citation bias, and citation noise using the example of a fictitious social citation system. This small world example which we use for illustration purposes in this study consists of citing and cited papers (see Table 1). We assume that the fictitious system includes all citing and cited papers of the small world: ten citing (1 to 10) and five cited papers (A to E). No other citing and cited papers exist outside the system: The system is an isolated system with no knowledge spillover or knowledge absorption from the external world. The citation decisions in the social citation system are from three citing authors (I, II, and III) for at least two different papers each. In Table 1, the citing authors are indexed with $i$, the citing papers are indexed with $j$, and the cited papers are indexed with $k$.

The binary values 0 and 1 are used in the table to indicate whether a (cited) paper is cited (column R), should be cited (column A) or is erroneously cited (column E) by a citing paper. For example, while citing paper 2 has cited (cited) paper D, citing paper 3 has not cited this paper. The author of citing paper 2 therefore attributes a different value to cited paper D than the author of citing paper 3. The values R=0, A=1, and E=1 for cited paper A (columns) and citing paper 1 (row) mean that author I has not realized to cite paper A in citing paper 1 although it should have. It follows that this citation decision of author I can be denoted as erroneous. Regarding the citation decisions made by the authors, we generally assume that citation decisions are reliable and interchangeable: Different authors would cite the same work at the same (or similar) point in a certain paper. That means citation decisions should not be based on personal taste or opinion but on objective requirements. Citation decisions may be trade-offs between the pros and cons of different citation options, but ultimately these trade-offs should be resolved by evaluative citation judgments made by the citing author about the perceived knowledge flow. It is possible, for example, that citing authors have found a certain fact to which they would like to refer in their own papers in several papers by another author. In this case, however, it is the task of the citing authors to make an evaluative citation judgment against the background of the available alternatives and the perceived knowledge flow. Large disagreements in performed citation decisions of different authors based on the same pool of available papers that are cited for one and the same text passage would violate expectations of reliability and validity in citation decisions (and citation analysis).

We assume in general and specifically for the social citation system in Table 1 that accurate citation decisions are possible for citing papers. It should be possible for authors to determine whether knowledge flow has happened or not. The assumed accurate citation decisions are given for each cited paper in the table (columns A). In a perfect citation world, all realized citation decisions (columns R) would be flawless measuring instruments of knowledge flow, and aggregated citations would be an error-free metric of knowledge flow. If Table 1 were free of errors, all realized citation decisions (columns R) would correspond to accurate citation decisions (columns A). For example, if cited paper A had always been cited accurately in the small world, all citing papers would have cited the paper. Since accurate citation decisions can mean that a certain publication should be cited (value 1) or should not be cited (value 0) in a paper, the columns in Table 1 with the accurate citations generally consist of patterns including 0 and 1 values.

In the social citation system, we see realized citation decisions as a measuring instrument for knowledge flow, where the decision instrument is the citing author. The citing author is confronted with a cloud of citation possibilities and makes an evaluative judgment (i.e. an assessment based on scientific criteria with respect to knowledge flow) about whether a particular work should be cited in the text or not. This human measuring instrument sometimes works better and sometimes worse and makes sometimes accurate and sometimes erroneous citation decisions. For the accurate measurement of knowledge flow in research evaluation processes, however, it is essential to avoid citation errors as far as possible and to increase citation decisions' reliability and accuracy. The aim should be to make citation decisions as accurate as possible.

We assume that a paper must be cited precisely when a text passage in the manuscript has a certain property: It is based on knowledge that is described in another paper. The citations in the list of references at the end of the manuscript should reveal thus the extent to which the present contribution (substantially) builds on previous contributions. The text passages in a manuscript where a certain other work should be cited as the proper knowledge context and the pool of works that could potentially be cited result in an expected pattern of citations for a given manuscript, which can be compared with the realized citations in the manuscript (see columns A and R in Table 1). The expected pattern can be used to determine for each manuscript the extent to which citation errors (columns E) are present as deviations from accurate citation decisions. These deviations can be determined and examined for cited and citing papers in a social citation system.

From the perspective of evaluative citation analysis, citation accuracy is the condition under which citation decisions correspond to knowledge flow. In practice, this condition is unlikely to be fully met, as the many studies on reasons for citing and inaccurate citations suggest. It is therefore reasonable to expect some degree of error – whether bias or noise – in the citation decisions associated with a manuscript. In the social citation system in Table 1, individual errors among citation decisions are denoted as $x_{ij}$. $x_{ij}$ denotes an erroneous citation made by citing author $i$ in citing paper $j$. Since we assume for the small world in Table 1 that all citing papers should cite a certain set of papers (A to E), we can use the binary data in the table to determine the degree of citation accuracy for each cited and citing paper. We start with the citing authors and focus on the lines in Table 1. The (fictitious) authors of the citing papers evaluated whether they should cite the five papers (A to E) and made accurate or erroneous citation decisions. For example, the author of citing paper 1 (author I) made a citation decision on five papers and made one accurate decision.

Binary data (typically 0 and 1) follows a Bernoulli distribution, where each observation has two possible outcomes. Given $n$ observations, where there are $a$ ones and $n$ - $a$ zeros, the mean $\mu$ is calculated as

$$\mu = \frac{a}{n} \qquad (1)$$

This means $\mu$ corresponds to the proportion $p$ that a value is 1. Since one of the five citation decisions for citing paper 1 in Table 1 is correct, the citing paper's proportion of accurate citations across the cited papers is $PA_{ij}$ = 0.2 (1/5) and the corresponding citing paper's proportion of errors is $PE_{ij}$ = 0.8 (4/5).

Each citation decision of the citing authors in Table 1 can be either:

- Correct positive: The citing author correctly identifies a paper that should be cited.

- Correct negative: The citing author correctly identifies a paper that should not be cited.
- Incorrect positive/negative: The citing author either wrongly includes a citation or fails to identify a necessary citation.

Citation accuracy for the citing papers ($PA_{ij}$) provides a snapshot of the citation decisions' quality by the citing authors considered in Table 1. As the results for the citing papers 1 to 10 show, $PA_{ij}$ is between 0.2 and 0.8. The average across all citing papers is $\overline{PA}$ = 0.54, which is only slightly better than a random selection of the papers to be cited by the authors. Given the prevalence of A and R in Table 1, the expected proportion of correct citations is about 49.8% if the authors randomly cite (i.e. irrespective of whether a citation is appropriate). The results indicate that the small world in Table 1 is affected by citation errors ($x_{ij}$) to a greater extent. It is therefore questionable whether its citation accuracy is sufficient to justify using the data in research evaluation processes. The results in Table 1 reveal thus that the validity of citation analyses used in research evaluation processes would benefit from determining the accuracy of citation decisions.

In a social citation system, we can also deal with accuracy at the level of cited papers. This involves the columns in Table 1. The paper citation accuracy is the proportion of correct citation decisions, which is calculated using the citation decisions of the ten citing papers (authors). The table indicates for each cited paper whether it should be cited by a citing author or not (column A). In the case of cited paper A, for example, we assume that it should have been cited by all 10 citing papers. As the values in Table 1 show, 6 author decisions that led to the realized citations are accurate; the cited paper's proportion of accurate citations is therefore $PA_k$ = 0.6. The proportion ($PR_k$) and times cited ($TC_k$) of the realized citations are aggregated values of citations for the individual cited papers A to E in Table 1. As Table 1 reveals for cited paper A, $PR_k$ = 0.6 and $TC_k$ equals 6. These values are the aggregated citation impact values that paper A has achieved in the small world under the influence of errors in the social citation system.

The various deviations of realized citations from accurate citations create noise in the social citation system in Table 1. Since deviations between realized and expected (accurate) citations differ for citing authors and citing papers within the system, variability emerges. We refer to this variability as noise. Noise is measured in the table using the standard deviation that is the most common measure of variability in statistics. The standard deviation σ is defined by

$$\sigma = \sqrt{\overline{PA} \times (1 - \overline{PA})} \quad (2)$$

The standard deviation represents the variance in the data with *n* observations, which is calculated as the square root of the proportion of accurate citations minus the proportion of erroneous citations. Citation noise is due to the variation among citing authors in their disposition to set citations in manuscripts.

Table 1. Fictitious social citation system including citing and cited papers

| | Cited paper A | | | Cited paper B | | | Cited paper C | | | Cited paper D | | | Cited paper E | | | | | | | |
|---|---|---|---|---|---|---|---|---|---|---|---|---|---|---|---|---|---|---|---|---|
| | $R$ | $A$ | $E$ | $R$ | $A$ | $E$ | $R$ | $A$ | $E$ | $R$ | $A$ | $E$ | $R$ | $A$ | $E$ | $PR_{ij}$ | $PA_{ij}$ | $PE_{ij}$ | $\overline{PE}_i$ | $\sigma_{PN,i}$ |
| Author I: Citing paper 1 | 0 | 1 | 1 | 1 | 1 | 0 | 1 | 0 | 1 | 0 | 1 | 1 | 0 | 1 | 1 | 0.40 | 0.20 | 0.80 | | |
| Author I: Citing paper 2 | 1 | 1 | 0 | 0 | 1 | 1 | 0 | 0 | 0 | 1 | 1 | 0 | 1 | 1 | 0 | 0.60 | 0.80 | 0.20 | 0.53 | 0.25 |
| Author I: Citing paper 3 | 0 | 1 | 1 | 1 | 0 | 1 | 0 | 0 | 0 | 0 | 0 | 0 | 0 | 1 | 1 | 0.20 | 0.40 | 0.60 | | |
| Author II: Citing paper 4 | 1 | 1 | 0 | 1 | 0 | 1 | 0 | 0 | 0 | 1 | 0 | 1 | 0 | 1 | 1 | 0.60 | 0.40 | 0.60 | 0.50 | 0.10 |
| Author II: Citing paper 5 | 1 | 1 | 0 | 1 | 0 | 1 | 1 | 1 | 0 | 0 | 0 | 0 | 0 | 1 | 1 | 0.60 | 0.60 | 0.40 | | |
| Author III: Citing paper 6 | 0 | 1 | 1 | 1 | 0 | 1 | 0 | 0 | 0 | 1 | 0 | 1 | 0 | 0 | 0 | 0.40 | 0.40 | 0.60 | | |
| Author III: Citing paper 7 | 1 | 1 | 0 | 1 | 0 | 1 | 1 | 0 | 1 | 1 | 1 | 0 | 0 | 0 | 0 | 0.80 | 0.60 | 0.40 | | |
| Author III: Citing paper 8 | 1 | 1 | 0 | 1 | 1 | 0 | 1 | 1 | 0 | 0 | 1 | 1 | 0 | 0 | 0 | 0.60 | 0.80 | 0.20 | 0.40 | 0.13 |
| Author III: Citing paper 9 | 0 | 1 | 1 | 0 | 1 | 1 | 0 | 0 | 0 | 0 | 0 | 0 | 0 | 0 | 0 | 0.00 | 0.60 | 0.40 | | |
| Author III: Citing paper 10 | 1 | 1 | 0 | 0 | 1 | 1 | 1 | 0 | 1 | 0 | 0 | 0 | 0 | 0 | 0 | 0.40 | 0.60 | 0.40 | | |
| $PR_k$ | 0.6 | | | 0.70 | | | 0.50 | | | 0.40 | | | 0.10 | | | | | | | |
| $TC_k$ | 6 | | | 7 | | | 5 | | | 4 | | | 1 | | | | | | | |
| $EC_k$ | | 10 | | | 5 | | | 2 | | | 4 | | | 5 | | | | | | |
| $PA_k$ | | | 0.60 | | | 0.20 | | | 0.70 | | | 0.60 | | | 0.60 | | | | | |
| $PE_k$ | | | 0.40 | | | 0.80 | | | 0.30 | | | 0.40 | | | 0.40 | | | | | |
| $\overline{PA}$ | | | | | | | | | | | | | | | | | 0.54 | | | |
| $\overline{PE}$ | | | | | | | | | | | | | | | | | | 0.46 | | |
| $\sigma_{LN}$ | | | | | | | | | | | | | | | | | | | 0.06 | |
| $\sigma_{PN}$ | | | | | | | | | | | | | | | | | | | | 0.17 |

Notes. $R$ = realized citation, $A$ = accurate citation, $E$= erroneous citation decision, $PR_{ij}$ = proportion of realized citations of citing author $i$ in citing paper $j$, $PA_{ij}$ = proportion of accurate citations of citing author $i$ in citing paper $j$, $PE_{ij}$ = proportion of erroneous citations of citing author $i$ in citing paper $j$, $\overline{PE}_i$ = Person-specific average error rate of citing author $i$, $\sigma_{PN,i}$ = author-specific pattern noise (within-author variance of erroneous citations for author $i$), $PR_k$ = cited paper $k$'s proportion of realized citations, $TC_k$ = times cited, $EC_k$ = expected citations, $PA_k$ = cited paper $k$'s proportion of accurate citations, $PE_k$ = cited paper $k$'s proportion of erroneous citations, $\overline{PA}$ = Proportion of accurate citations in the entire citation system, $\overline{PE}$ = Proportion of erroneous citations in the entire citation system, $\sigma_{LN}$ = overall level noise (between-author variance of erroneous citations), $\sigma_{PN}$ = overall pattern noise.

In the social citation system, there are three different types of citation noise: citation level noise, citation pattern noise, and citation occasion noise. These three types of noise form the overall citation noise in the social citation system. What do we mean by citation level noise? Some citing authors are more prone to make erroneous citations than others. Some authors tend to cite too little; others tend to cite too much. $\overline{PE_i}$ gives the proportion of erroneous citations across all an author's citing papers and can thus be interpreted as the author-specific citation error rate. For example, Table 1 shows that citing author I has an error proportion of approximately 0.53, which is higher than the corresponding proportion of 0.40 for citing author III. These author specific level differences in the social citation system primarily reflect personal citation styles, which may also be shaped by subject-area conventions. From the perspective of evaluative citation analysis, citations would ideally follow a shared interpretation – a paper is cited when knowledge flow from the cited to the citing work has occurred. Actual citation practices, however, vary across authors and conventions. We refer to this variability in erroneous citation decisions across citing authors as citation-level noise.

In statistics, this type of variability is referred to as between-author variance. It captures the variance of erroneous citation decisions across different citing authors. Citation level noise ($\sigma_{LN}$) is measured as the standard deviation of the author-specific error rates ($\overline{PE_i}$) from the proportion of erroneous citations in the entire citation system ($\overline{PE}$) weighted by the number of citing papers co-authored by citing author $i$ ($n_i$).

$$\sigma_{LN} = \sqrt{\frac{\sum_i n_i(\overline{PE} - \overline{PE_i})^2}{\sum_i n_i}} \quad (3)$$

Using the values in Table 1, the citation level noise in the fictitious social citation system is $\sigma_{LN}$ = 0.06. The small amount of citation level noise indicates that the citing authors have a similar level of citation accuracy.

We now turn to citation pattern noise: There is more noise in the system than can be explained by different error levels of the citing authors. Whereas citation level noise refers to (erroneous) inconsistencies in the citation decisions of different authors, citation pattern noise refers to inconsistencies in the citation decisions of one and the same author. More specifically, citation pattern noise is about the interaction of citing author characteristics and cited paper characteristics. In Table 1, the citation pattern noise values for each citing author can be found in column $\sigma_{PN,i}$: It is the within-variance of citation errors for each citing author.

$$\sigma_{PN,i} = \sqrt{\frac{1}{n_i}\sum_{j=1}^{n_i}(\overline{PE}_i - PE_{ij})^2} \quad (4)$$

The within-author variance of citation errors is calculated as the standard deviation of citation errors in an author's citing papers ($PE_{ij}$) from the author-specific citation error rate ($\overline{PE}_i$). $n_i$ denotes the number of citing papers authored by a specific citing author $i$. As the results in Table 1 show, the (citing author) pattern noise is similar for all authors: It is between $\sigma_{PN,i}$ = 0.10 and $\sigma_{PN,i}$ = 0.25.

$$\sigma_{PN} = \sqrt{\frac{\sum_{i=1}^{n}(n_i \times \sigma_{PN,i})}{\sum_{i=1}^{n} n_i}} \quad (5)$$

The overall citation pattern noise in the social citation system ($\sigma_{PN}$) is defined as the weighted average of the author-specific citation pattern noise values, whereby $n$ denotes the number of citing authors. The results in Table 1 reveal that the overall citation pattern noise amounts to $\sigma_{PN} = 0.17$.

Citation pattern noise consists of two components: (1) stable citation pattern noise and (2) occasion noise – the random part of pattern noise. Stable citation pattern errors are assumed to be unique to each citing author: Even if a citing author will generally tend to make more or fewer erroneous citations, this tendency will furthermore be determined by certain properties of the papers that are available for citation decisions. A citing author might, for example, tend to cursory or careless citations in writing, but is more inclined to erroneously cite papers that propose specific statistical methods (the citing author is not a statistician) than papers in his subject area (he has a keen interest in these papers). Another author tends to erroneously cite only a few papers; however, the author frequently makes citation errors in methods papers because she doesn't understand them. Such interactions of citing and cited papers properties vary from citing author to citing author. Stable citation pattern noise is not a random event in the citation process, but rather an error interaction between citing author and cited papers that occurs systematically.

Pattern noise can not only occur systematically but can also be attributed to random events. Gigerenzer and Brighton (2009) speak of "accidental patterns" (p. 120). We would like to refer to these (erroneous) random citation decisions as citation occasion noise (which is also pattern noise, but not stable pattern noise). We have described such random citation decisions in the introduction of this paper by quoting Chen et al. (2024). Citation occasion noise arises, for example, when the writing author Z of a paper Z accidentally becomes aware of a text passage in another paper Y. This passage is taken up by author Z and inserted as a citation in paper Z, although author Z mistakenly interpreted the passage as highly suitable for paper Z by reading paper Y quickly and superficially. At another point in time, author Z would probably not have cited paper Y in paper Z (but another paper or no paper at all). Author X tends to superficially read a lot of papers by colleagues and therefore erroneously cite many papers in paper X (and many other papers). However, as the author was on vacation and had plenty of time to read the papers written by colleagues for his planned paper Q, the author cites less erroneously in Q. This within-citing author variability, which can be determined via test-retest reliability, is conceptually different from the stable between-citing author differences: we call this variability based on transient effects citation occasion noise. We do not provide a mathematical formula for measuring citation occasion noise. Kahneman et al. (2021) discuss it conceptually, not mathematically.

We have now described all types of noise that can occur in social citation systems according to Kahneman et al. (2021). Citation system noise is the undesirable variability in citation decisions on the available papers in a social citation system that can be potentially cited by multiple citing authors. Citation system noise includes two noise components, which can be separated when the same citing authors evaluate multiple papers that can potentially be cited: (1) Citation level noise is the variability in the author-specific tendency to make citation errors. The amount of citation level noise in a social citation system is small if all citing authors have a similar tendency to make erroneous citations. By contrast, there is a large amount of citation level noise if some authors make substantially more accurate citation decisions than others. (2) Citation pattern noise is the variability in citing authors' responses to the characteristics of papers that can potentially be cited. Citation pattern noise is large if citing authors are very inconsistent in their citation errors and it is small if the citing authors are very consistent in their tendency to make erroneous citations. (2a) Stable citation pattern noise arises in the interaction between the characteristics of the citing author and the characteristics of the papers that they could cite. This pattern noise is rooted in the stable personal characteristics of the citing authors regarding his or her erroneous citation decisions. (2b) In addition to stable citation pattern noise, there is also citation occasion noise, which can be described as random error. Citation level noise (1) and citation pattern noise (2a, stable pattern and, 2b, occasion noise) both contribute to the overall system noise.

The overall citation system noise is thus the combination of citation level noise and citation pattern noise. Citation system noise is the total amount of citation noise in the social citation system

$$\sigma_{SYS} = \sqrt{(\sigma_{LN})^2 + (\sigma_{PN})^2} \quad (6)$$

We used the values $\sigma_{LN}$ = 0.06 and $\sigma_{PN}$ = 0.17 in Table 1 to compute the total amount of citation noise in the (fictitious) social citation system. The resulting $\sigma_{SYS}$ = 0.18 indicates a certain level of noise in the system.

In addition to the various types of noise that can be identified in the social citation system, we also must deal with various citation biases that contribute to erroneous citations. Citation bias refers to a systematic, directional error in citation decisions. If several citing authors in Table 1 systematically exhibit citation errors due to certain factors (which are independent of knowledge flow and have a causal effect), we speak of citation bias. Citation decisions are biased if the cited papers are cited (significantly) more often or less often than what would be expected if all citation decisions were accurate. In other words, citation bias is measured by the difference between the average number of realized citations $(TC_k)$ and the average number of expected (accurate) citations $(EC_k)$ in a set of cited papers:

$$\frac{1}{n_k}\sum_{k=1}^{n_k} TC_k \neq \frac{1}{n_k}\sum_{k=1}^{n_k} EC_k \quad (7)$$

In the formula, $n_k$ denotes the number of cited papers in the social citation system. In Table 1 for example, the cited papers should have received an average of 5.2 citations if all citations decisions were accurate [(10 + 5 + 2 + 4 + 5) / 5]. However, the papers actually received 4.6 citations on average [(6 + 7 + 5 + 4 + 1) / 5]. Therefore, there is a downward citation bias in the social citation system in Table 1: on average, the cited papers are under-cited.

Citation biases arise in a social citation system when many errors in the system have a specific tendency that relates to characteristics of the people (authors), institutions, countries, journals, etc. involved in the system (i.e. causally irritating the system). As we mentioned above, Jannot et al. (2013), for example, found that there is a citation bias favoring statistically significant results in medical research. On average, therefore, more citations are (erroneously) attributed to papers with statistically significant results than to papers without these results that could have been cited equally. Mutz et al. (2017) were able to show that papers that were labeled as 'very important paper' by a reputable journal received more citations than comparable papers (i.e. papers of similar quality) without this label. Accordingly, papers labeled as very important receive on average more (erroneous) citations than papers without this label, i.e. citations that are not covered by a corresponding knowledge flow from the cited to the citing paper.

Unlike citation noise, biases in citation decisions have already been dealt with in many ways in bibliometric research (Kousha & Thelwall, 2024). Although we can assume that citation bias plays at least as great a role as citation noise in citation decisions, the two phenomena are treated very differently in bibliometric research. The difference in treatment is probably because – on the one hand – researchers tend to explain phenomena, and we can explain biases by linking them to a specific triggering factor. Researchers therefore try to assign a reason or factor to certain citation decisions, such as papers labeled as 'very important paper'. Biases in citation decisions can relate to characteristics of the citing author (such as their school of thought) or characteristics of the cited paper (such as the school of thought of the cited authors). Citation noise – on the other hand – is a statistical phenomenon that can hardly be linked to specific characteristics: noise can only be identified if the distribution of an ensemble of citation decisions is statistically analyzed.

# 3 Strategies to reduce citation noise and citation bias

Over the past decade, several influential reports and initiatives – a "professional reform movement" (Rushforth & Hammarfelt, 2023, p. 879) – have stressed the need for careful, responsible use of bibliometric indicators in research evaluation. Key examples include the Leiden Manifesto (Hicks et al., 2015), the United Kingdom (UK) Metric Tide report (Wilsdon et al., 2015), and a European Union (EU) agreement on research assessment reform (Coalition for Advancing Research Assessment, CoARA, see https://coara.eu). In parallel, agreements such as the Declaration on Research Assessment (DORA, see https://sfdora.org/) have pushed to limit the reliance on journal-based impact metrics, while the More Than Our Rank initiative (see https://inorms.net/more-than-our-rank/) highlights that university rankings – which often incorporate citation-based indicators – fail to capture essential aspects of an institution's performance (Thelwall, 2024). Collectively, this reform movement underscores the limitations of depending solely on bibliometric measures in research evaluation. According to Thelwall et al. (2023), the reform movement delivers "strong arguments against using citations for aspects of research assessment" (p. 941).

The reform movement has drawn attention to the risks of using citation metrics in research assessment (Rushforth & Hammarfelt, 2023). The debate treats citation metrics mainly as a problem of evaluation systems and not as a problem of citation practices that generate the data on which these systems rely. This distinction matters because citations occupy a dual role: They are not only used as data in research evaluation processes, but also – and that is often forgotten – a basic practice of scholarly communication. Researchers are expected to cite prior work to show what their own study builds on, and to make their arguments traceable. In this sense, accurate citations are treated as a meaningful scholarly practice. This is underscored by findings from a survey of psychology-journal editorial board members conducted by Lawson et al. (2025): "93% of respondents agreed that 'citing is foundational to the credibility of psychological scholarship'" (p. 2).

If accurate citation is regarded as an essential part of good research practice, it is difficult to maintain that citations have little or no relevance for research evaluation. The central issue of a reform movement should not be the use of citation metrics per se, but the quality and standardization of citation practices from which citation metrics are derived. When citation decisions are inconsistent or shaped by factors unrelated to knowledge flow, citation indicators become less valid. When citation practices are more rigorous and standardized, citation data become more informative for evaluative purposes. Improving citation analysis requires attention not only to the responsible use of indicators, but also to the scholarly practices that produce citation data. The consideration of citation decisions underlying citation-based indicators in efforts to reform is especially relevant for initiatives that address the responsible use of bibliometric indicators. Citation analysis can provide valid information about research impact only to the extent that citation decisions accurately reflect knowledge flow from cited to citing work. Improving citation practices should therefore be seen as part of responsible research assessment, not as separate from it. According to Penders (2018), "citation-based metrics have proliferated as proxies for quality and impact over the years, only to be currently subjected to significant and highly relevant critique. To cite well, or to reference responsibly, is thus a matter of concern to all scientists".

Before we move on to strategies for reducing citation noise (and bias) in the social citation system which can be taken up by the reform movement, we would first like to raise awareness of the topic of citation noise in the context of patent citations, as citation noise has already played a major role there (including strategies for reducing citation noise). Patent citations should represent (intensity of) knowledge flow. Cited and citing patents are seen as proxies of knowledge flow; cited references represent how previous knowledge has been combined in citing patents to produce new knowledge. Noise identification in the knowledge flow between patents is primarily about the distinction between relevant and irrelevant prior art citations. Smith (2014) describes the emergence of citation noise in patents as follows: "Citation noise occurs due to timing in the patenting process, the point in time when the prior art references are identified and the amended state of the patent claims at that point in time. Citation noise also occurs due to different perspectives and comprehension concerning the technology and invention. Prior art references may be provided by many actors such as the inventor with the patent application, a searcher during a patent office search phase, an examiner during a patent examination phase, or an interested third party after publication of the application. Inventor-supplied citations are typically addressed before filing a patent application and are typically not relevant at the time of filing the patent application. Searcher-supplied citations and third party supplied citations may arrive later, after amendments. They may not be relevant due to the preceding amendments that redefine the invention" (pp. 36-37).

According to Smith (2014), current citation-based patent evaluation methods are constrained by citation noise, which may lead to insufficient indicators of the economic value of patents and may obscure thus valuable economic insights. We can assume that we have a similar problem with evaluative citation impact analyses in bibliometrics – if half of the citations are of the perfunctory and persuasive type (Bornmann & Daniel, 2008) and 10% to 20% of citations are inaccurate (Wakeling et al., 2025). By focusing on relevant (accurate) patent citations, noise may be eliminated, uncovering hidden connections by identifying, for example, strategic relationships and enhancing technology and invention assessments. A patent may include backward-pointing citations that refer to earlier prior art, and it might also serve as a forward citation for later patents. Combined, these two types of links establish a network of meaningful relationships between patents and their citations – provided they can be discerned amid the background noise of patent citation data.

Just as Smith (2014) points out the importance of accurate citations in patent analysis, ensuring citation accuracy is also crucial in bibliometric citation analysis. The goal should be the reduction of error in citation data, which refers to both bias and noise. More than ten years ago, Murphy (2011) already pointed out that "the emergence of PubMed has resulted in a citation explosion and a resulting quagmire as many authors are just not selective about referencing papers and tend to be a bit over zealous in citing more and more papers many of which are not on the topic" (p. 307). The author was troubled by the growing trend in which authors cite papers they have not thoroughly read. His concern is that many of these authors may have engaged with nothing beyond a paper's title. The authors' decision to cite a work is not based on a comprehensive understanding, but rather on a superficial impression gleaned solely from the title or abstract. These citation decisions may be an explanation therefore for why citation errors occur. However, the citation error medal not only has the side of careless and superficial citation, but also the side of plagiarism by authors. In plagiarism, authors take content from another publication, but this knowledge flow is not indicated by a citation (Masic, 2013). The authors who have made a knowledge flow fail to insert a citation in their text, although they should have done so – in the case of an accurate citation decision.

In the following, we would like to present some methods with which citation errors (noise) can be reduced and the proportion of accurate citations can be increased. On the one hand, this involves the method of aggregation of citation decisions, which is common in bibliometrics, and on the other hand, measures for improving citation decision hygiene such as guidelines and training for accurate citation decisions. These measures are generally aimed at reducing citation noise in general, i.e., without focusing on specific citation errors.

## 3.1 Aggregation of citation decisions

Aggregating citations in citation analysis builds on what Galton (1907) termed the 'wisdom of crowds' at the dawn of the twentieth century. The author reports on a contest held at an English livestock show, where 787 participants estimated the weight of a publicly displayed ox. Remarkably, the median estimate of 1,207 pounds was within less than 1% of the ox's actual weight of 1,198 pounds. Estimating the ox's weight by averaging over a larger number of independent estimates was probably one of the first demonstrations of the advantage of the crowd over a single opinion: When a sizable group offers judgments or estimates, the combined result tends to be highly accurate. The insight of Galton (1907) can also be applied to citation analysis. Citations can be seen as reflecting the evaluations of many researchers. Because nearly all researchers disseminate their research findings through publications, and cite prior work in doing so, citation data capture the judgments of an enormously large community of citing researchers. Eugen Garfield – the founder of the Science Citation Index – "called his Science Citation Index an 'association-of-ideas index'. He relied on the judgment of experts, members of the research community itself, whom he called his 'army of indexers'" (Pendlebury, 2025).

We turn to the fictitious social citation system we present in section 2 to explain in detail how and why citation analysis profits from the aggregation of citation decisions. When each author's citation decision is modeled as a binary outcome (with a probability $p$ of citing a paper), the variance of that decision is given by $p(1 - p)$. We have already dealt with the calculation of the variance in section 2. The average (or normalized sum) of many such binary variables (with the number $n$) forms the basis of a binomial distribution, and – by the central limit theorem – the standard error of the average decreases proportionally to $1/\sqrt{n}$. Two (extreme) examples (Table 2 and Table 3 in the appendix) should illustrate the relationship between citation bias and citation noise, on the one hand, and aggregated citation counts, on the other hand. The social citation systems in Table 2 and Table 3 deviate slightly from the scenario in Table 1. Table 1 contains five cited papers and 10 citing papers. Table 2 contains only four citing papers and Table 3 contains 11 citing papers. These changes were made to illustrate the two ways in which incorrectly given and incorrectly missing citations can average out on the aggregate level.

The first example in Table 2 is a fictitious social citation system, in which each cited paper is either over-cited ($TC_k > EC_k$) or under-cited ($TC_k < EC_k$), despite the absence of citation bias and citation noise. The second example in Table 3 is another fictitious social citation system, in which every cited paper has been accurately cited ($TC_k = EC_k$) even though there is an extreme amount of citation noise in the system. In both examples, the citations are unbiased because incorrectly given and incorrectly missing citations cancel out when the citations are aggregated across cited papers. The examples demonstrate that the larger the number of cited papers is, the larger the improvement in citation accuracy gained by aggregating citation decisions. However, aggregating citation decisions across numerous cited papers does not ensure accuracy on the level of individual cited papers.

The accuracy of an individual cited paper's citation counts strongly depends on the amount of citation noise in the system and on the number of independent citation decisions made by potentially citing authors. The greater the number of independent citation decisions per cited paper, the higher the likelihood that incorrectly given and incorrectly missing citations cancel out on the individual paper level. We emphasize the need to distinguish between realized citations and citation decisions. In citation analysis, the sum of realized citations is typically interpreted as a paper's impact. However, the accuracy of citation analysis does not technically stem from the aggregation of realized citations, but from the aggregation of citation decisions, which may or may not result in realized citations. Even when a publication receives no citations, a citation count of zero can be considered a meaningful aggregate, provided that the decision not to cite the publication was made independently by a sufficiently large number of researchers.

The assumption that researchers make citation decisions independently of the citation decisions of others warrants closer examination. So far, we have focused on the benefits of aggregating citation decisions; however, not all crowds make sound decisions. Surowiecki (2004) emphasizes that collective judgments achieve high accuracy only when individual opinions are formed independently. The specific type of independence required for sound collective judgements is stochastic independence, meaning that the probability of one researcher citing a paper is not affected by whether another researcher has cited it. Because a paper's citation count is publicly available information (for example, displayed on journal websites), it seems likely that researchers are aware of – and influenced by – the citation decisions of others. Hypotheses concerning the non-independence of citation decisions form a central foundation of citation theory. Price (1976) proposes a general theory of cumulative advantage processes (sometimes referred to as preferential attachment) in citation distributions. Cumulative advantage in citation decisions means that "the probability that an author chooses a particular paper to cite is not uniform, but proportional to how many citations the paper already has" (Wang & Barabási, 2021, p. 185). Several studies documenting the high concentration of citations at the top of the citation distribution discuss cumulative advantage as a mechanism that contributes to the highly unequal distribution of citations across papers (Dougherty et al., 2024; Kim et al., 2020; Nielsen & Andersen, 2021; Varga, 2022). The non-independence of citation decisions implies that citation errors are distributed non-randomly across cited papers, which in turn generates biases that do not average out when citation decisions are aggregated (unlike noise). Cumulative advantage and other mechanisms of non-independence therefore pose a serious challenge for citation analysis that cannot be resolved simply by using large datasets.

In summary, aggregating multiple independent judgments harnesses the 'wisdom of crowds' by averaging individual assessments to diminish noise. Citation analysis profits from two distinct types of aggregation depending on the level of the analysis: Aggregating citation counts across cited papers helps to combat the undesirable consequences of citation noise in sufficiently large samples. On the level of individual cited papers, the same is achieved by aggregating independent citation decisions. The amount of aggregation required to ensure the reliability of citation counts depends on the amount of citation system noise: The larger the amount of citation noise in the system, the more aggregation is required to perform valid and reliable citation analysis. Although not dealing with citation noise in citation decisions, van Raan (2005) argues for the method of aggregation in targeting citation errors: "So undoubtedly the process of citation is a complex one, and it certainly not provides an 'ideal' monitor on scientific performance. This is particularly the case at a statistically low aggregation level, for example the individual researcher. There is, however, sufficient evidence that these reference motives are not so different or 'randomly given' to such an extent that the phenomenon of citation would lose its role as a reliable measure of impact. Therefore, application of citation analysis to the entire work, the 'oeuvre' of *a group of researchers as a whole over a longer period of time* (author's emphasis), does yield in many situations a strong indicator of scientific performance" (pp. 134-135).

The research community is certainly a large crowd, but it is not a crowd that makes only independent and unbiased decisions. Kahneman et al. (2021) point out that, in general, aggregating decisions may reduce noise but does not necessarily reduce the bias stemming from individual judgments. We suspect that the lack of reduction of biases in aggregated citation decisions is one reason why citation biases have been investigated very frequently in bibliometrics to date – in contrast to citation noise. The extent to which citation decisions are biased by non-independence could certainly be reduced if citing authors were made more aware of the issues of citation accuracy, citation bias, and citation noise.

### 3.2 Citation decision hygiene

In addition to the aggregation of citations, another strategy for improving the informative value of citation analyses is various measures to improve citation decision hygiene, which are presented in this section. The survey of psychology-journal editorial board members conducted by Lawson et al. (2025) shows that “86% of editorial-board members at least slightly agreed that ‘thus far, our field has not paid enough attention to improving the quality of citing decisions’” (p. 3).

#### 3.2.1 Guidelines for ensuring accurate citations

If valid citation analysis requires citation decisions to be sufficiently standardized, then citation guidelines are one possible way to support such standardization. Citation guidelines could help authors make citation decisions more consistently during the writing process. Their purpose would not be to prescribe citation behavior in general, but to clarify the conditions under which citations are most informative for citation analysis. From this perspective, citation decisions should ideally be based on identifiable knowledge flow rather than on superficial reading, individual preference, or situational factors. Knowledge flow occurs when elements of a manuscript draw on prior published work. These elements may include, for example, an empirical finding, a theoretical concept, a methodological approach, a dataset, or an argument introduced in an earlier publication. Guidelines could help authors identify such cases and cite the specific works from which these knowledge elements were taken.

Superficial reading, individual preferences, and temporary moods are sources of citation noise. Although scholarly publishing relies on peer review and editorial oversight, many citations in a paper do not correspond to the content of a previous work or do not support the statement it accompanies (Wakeling et al., 2025). According to Lawson et al. (2025), “many researchers do not read, evaluate, and cite prior works as comprehensively as they would hope. As a result, current citing behaviors can increasingly be understood through the lens of ‘satisficing,’ or employing less effortful cognitive strategies that are minimally acceptable to accomplish tasks … Satisficing is contrasted with ‘maximizing,’ a process that involves expending more effort to weigh options and optimize a solution” (p. 8). In this paper, we argue not only that citation guidelines should be developed to improve citation decision hygiene at citing authors, but also that these authors should be encouraged to follow these guidelines through appropriate journal policies to guiding not only authors but also reviewers and editors. Reviewers and editors should be instrumental in detecting citation errors; systematic research into peer review practices for verifying citation accuracy is therefore important to know which guidelines are necessary (Wakeling et al., 2025).

Citation guidelines for researchers already exist, but these guidelines generally deal with the format and style of citations (for example, Harvard style or Chicago style). Following citation styles may reduce citation noise caused by cited reference errors in literature databases, but the guidelines do not provide any guidance on when it is necessary to cite a particular publication and when it is not (Klitzing et al., 2019). For example, the publication manual of the American Psychological Association (2020) recommends that the authors cite "the work of those individuals whose ideas, theories, or research have directly influenced" (p. 253) their work. However, Lawson et al. (2025) note that its guidance remains heavily focused on plagiarism and formatting: the manual contains "24 pages focused on plagiarism and citation formatting, less than half a page of the manual is dedicated to other aspects of citing, which may signal that authors have total license to make highly subjective decisions about whom and what to cite" (p. 3).

From the perspective developed here, this lack of explicit guidance matters because citation decisions that rely mainly on implicit assumptions, disciplinary habits, or intuitive judgments are likely to vary across authors. Such variation can contribute to citation noise. Kahneman and Frueh (2022) argue more generally that structuring decision-making processes is an important way to reduce noise in human judgment: "Try to design an approach to making a judgment or solving a problem, and don't just go into it trusting your intuition to give you the right answer" (p. 25). Applied to citation practices, this suggests that clearer guidance on what to cite could help standardize citation decisions and reduce reliance on highly subjective criteria. We assume that citation guidelines can be an efficient mechanism for the reduction of noise, because they specify exactly when a citation should be made, thereby reducing the variance between the authors' citation decisions.

The need for better guidance on correct citation behavior is illustrated by the study of Bruton et al. (2025), who conducted a survey on such questions and questions on ethically questionable citation behaviors among 257 U.S. researchers. Through the respondents' answers, the authors were able to group ethically questionable citation behaviors into three categories: strategic citations, neglectful citations, and blind citations. Furthermore, the authors found that ethically questionable citation behaviors do not depend on the length of the scientific career: all researchers are affected. For the authors, the empirical results suggest that citation norms should be more clearly articulated in academia and that there should be improved guidance about citations. Their study shows that "rarely are the ethical norms of citations articulated clearly and meticulously" (Bruton et al., 2025, p. 331).

To the best of our knowledge, one of the few guidelines about citing accurately has been published by Murphy (2011). Even if the rules of Murphy (2011) refer to different aspects in connection with citation decisions, they essentially concern the indication of knowledge flow by the citing author:

1. "Read and comprehend all of the literature that you are citing in your manuscript
2. Cite the primary literature and the actual papers to which a particular discovery is attributed; if multiple citations need to be made, do so
3. Cite the literature that agrees as well as that which disagrees with your point-of-view; be fair to multiple points-of-view
4. Be as inclusive as possible, but know where to draw the line, citing the most pertinent literature and the original papers
5. Use reviews judiciously in your manuscript and use them properly
6. Do NOT cite reviews in lieu of citing the original literature

7. Remember that PubMed [a literature database in biomedicine] goes back to 1966, but many worthwhile discoveries were made prior to that time and it is important to be historically accurate in your citations
8. When citing your own work, do so to support your point-of-view or to put the current work into the proper context for the reader. Do not attempt to grow your own citation base solely on the dreaded 'self-citation'" (p. 309).

In addition to this list of eight rules in the citation guidelines, the literature is otherwise rather fragmented when it comes to rules and necessities in connection with citation decisions. Penders (2018) points out, for example, that citations should be used to differentiate your own research from that of others, and that citations in papers can provide readers with references to other relevant literature: Omitting citations to relevant previous publications "can wrongfully suggest that your own publication is the origin of an idea, a question, a method, or a critique, thereby illegitimately appropriating them. Citations identify where ideas have come from, and consulting the cited works allows readers of your text to study them more closely, as well as to evaluate whether your use of them is appropriate" (Penders, 2018). In addition, the author opposes the use of superficial citations: "Ask yourself why you are citing prior work and which value you are attributing to it, and whether the answers to these questions are accessible to your readers" (Penders, 2018).

#### 3.2.2 Citation justification table

To improve the citation decision hygiene among authors, Lawson et al. (2025) propose that journals should request from authors "an annotated reference section with approximately one to two sentences for why each cited work was included in the manuscript" (p. 11). Building on this proposal, one possible measure to improve citation decisions would be to include a citation justification table in an appendix of research articles. In such a table, authors could briefly indicate why a particular publication was cited at a specific point in the manuscript. In the terminology used here, the table would document the knowledge flow associated with each citation. As an example of a citation justification table, we have generated a corresponding table in the appendix of this paper where we explain the reasons for every citation included in this paper (see Table 4). References to such citation justification tables can already be found at Penders (2018): "Sources deserve credit for the exact contribution they offer, not their contribution in general. This may mean that you need to cite a single source multiple times throughout your own argument, including explanations or indications why".

By creating a citation justification table, authors are incentivized to cite relevant publications (otherwise an author could not enter anything in the table). The table could also be used by reviewers of a paper to check the quality of the citation decisions in the paper: Reviewers can assess (1) whether publications are missing that should be cited, (2) whether cited publications may not have been read or understood, and (3) whether citations do not reflect the true content of the cited paper. In addition, the information in these tables on the papers could be used to classify citations in the papers ex post (by providers of literature databases such as Clarivate). In this classification, however, it would no longer be a question of whether the citation is superficial or substantial, but rather which knowledge elements were taken from prior literature, such as a statistical method used, an inspiring idea, a certain empirical result or an underlying theory. Thus, the meaningfulness of ex-post citation classifications would increase significantly.

#### 3.2.3 Training of (early-career) researchers

As citation decision hygiene has hardly played a role in the training of early career researchers to date (to the best of our knowledge), we would consider it necessary for (early career) researchers to be trained in this area in the future. Lawson et al. (2025) point out the problem of the lack of training in psychology as follows: "When citing, psychologists make a series of highly consequential decisions. Yet with little formal training or guidance on how to approach this process, many scholars implement ad hoc, informal, or effort-avoiding strategies that can undermine the cumulativeness and inclusiveness of the field" (p. 1). Young researchers should be trained in how to make accurate citations to the literature used in their own papers. In the survey of psychology-journal editorial board members conducted by Lawson et al. (2025) "most (84%) agreed that 'most graduate programs should do more to train students in how to approach citing decisions'" (p.3).

During training, young academics should be taught the difference between substantial and superficial citations. Positive and negative examples from scientific literature can be used for this purpose. The young researchers should learn about best citation practices in literature and what is meant by knowledge flow. As early career researchers often base their scientific work on what has been published in the publication manuals of professional associations (such as the publication manual of the American Psychological Association, 2020), these manuals should also include detailed rules to promote citation decision hygiene. Instructional videos on citation decision hygiene might be additionally helpful besides the instructions in publication manuals.

#### 3.2.4 Correction of citation errors

Another element in citation decision hygiene could be the identification and correction of citation errors in publications that have already been published. Since scientific literature is usually published digitally these days, it should be possible to correct errors in citations retrospectively. However, this process should not only be in the hands of the cited author but should also be supported by staff at the relevant publisher or platforms. For example, PubPeer (see https://www.PubPeer.com) is an online platform for post-publication peer review, where researchers can comment on published research to highlight errors, discuss methods, or flag possible misconduct. A correction of citations should only be possible if the cited author can clearly demonstrate that the correction to a citation is necessary. We were only able to find one study that has dealt with the possible correction of citations by cited authors in the past.

Wakeling et al. (2025) asked authors whether they had ever come across a citation of their work that they considered inappropriate and, if so, what measures they took. The results are as follows: "While 43.0% of respondents said they had never encountered an inappropriate citation of their work, 46.4% said they had, but had taken no action. 9.3% of respondents said they had contacted the authors of the citing article that inappropriately cited their work, while 3.1% said they had contacted the journal editor, and 0.7% the publisher" (Wakeling et al., 2025, p. 1404). The results reveal that only a few authors have so far considered the possible correction of their works' citations in other publications. Researchers are probably disincentivized from correcting erroneous citations of their work. If citations exist in the scholarly record, they count as citations in the various citation-related metrics that researchers are generally seeking to maximize.

#### 3.2.5 The use of AI for citation decision correction

As a final building block for the promotion of citation decision hygiene, we would like to mention the potential use of artificial intelligence (AI). In recent years, several application areas have been presented in the literature in which AI could be integrated into the work process of researchers. An overview of the various areas in which tasks can be performed with astonishing speed and accuracy formerly regarded as quintessentially human can be found at Binz et al. (2025) and Wang et al. (2023). One of these areas also involves suggesting scholarly references for a scientific text. Sourcely (see https://www.sourcely.net) is an AI-powered academic research assistant that helps you find, summarize, and cite scholarly sources by analyzing your text and matching it with relevant papers. Elicit (see https://elicit.com) also is an AI-powered research assistant that helps you search, summarize, screen, and extract data from scientific papers, automating many steps of finding relevant literature to cite. Paperpal (see https://paperpal.com) is an AI-powered academic writing platform that helps researchers read, write, edit, and cite more effectively.

Algaba et al. (2024) dealt with suggesting scholarly references for a scientific text in an experiment based on 166 papers. The authors used the large language model (LLM) GPT-4 from OpenAI to suggest scholarly references for anonymized scientific texts. Even though the authors observed "a remarkable similarity between human and LLM citation patterns" and stated that LLMs "can aid in citation generation", the authors found that the generated citations "may also amplify existing biases and introduce new ones, potentially skewing scientific discourse" (Algaba et al., 2024). In an earlier study by Khan et al. (2023) in the field of stem cell research, the authors similarly point out that "ChatGPT has the ability to produce generally accurate references, although it was observed to occasionally generate artificial hallucinations" (p. 5275).

In this study, we do not advocate using AI in the academic writing process to insert citations into a text if knowledge flow from a previous publication can be assumed for this text. Since only the author of a text can judge when a knowledge flow has taken place and when it has not, no AI can do this autonomously. However, no author is infallible in the writing process. We can therefore imagine AI being used in this process to identify text passages in a manuscript where a quote may have been forgotten by the author or where a citation may have been erroneously inserted. AI would therefore be used ex ante – before a scientific text is published – to check citation decisions of authors as a debias and denoise instrument. This instrument could be used by the authors themselves or the journal to which a manuscript was submitted. AI would mark places in the text where it might be necessary to cite another work. This would be the case, for example, if the citing author uses knowledge that has already been published, refers to a theory without having cited the underlying work, or the citing author forgets to cite the corresponding publication when using a method.

Following studies such as those presented by Khan et al. (2023) and Algaba et al. (2024), future experimental studies should examine whether AI can be used to inspect citation decisions of authors. If it turns out to be possible, it should be determined how citing authors can use AI to control their citations. We could imagine AI being used by researchers even before the writing process begins, for example to identify previous research that is relevant to their own research. What has already been empirically researched? Which methods should be used in the own research? Which theories could be relevant for this own research? Since AI can draw on a broad base of scientific literature, it may be very helpful in preparatory processes for the own research.

## 4 Discussion

Citations are a fundamental component of scientific communication, representing the flow of knowledge between research papers. Research papers include 'codified knowledge' "that is addressed to a large and partly anonymous audience" (Aman & Gläser, 2025, p. 158). Since the anonymous audience uses the codified knowledge as a basis for future research, citations of previous papers are integral components to the evaluation of research output and impact. Since the accuracy of the underlying citation data is paramount for the validity of such evaluations, this paper delves into the critical issues of citation noise and bias. Citation noise and bias can distort the 'true' representation of knowledge flow and, consequently, the reliable and valid assessment of research output and impact. Arbitrary citations contradict the meaningfulness of citations for research evaluation. With the conceptualization of citation accuracy, citation noise, and citation bias in a knowledge flow framework of citations, this study is intended to provide a foundation of citation analysis in bibliometrics. With our study, we are responding to calls like the one from Milojević (2025) "to invest effort not only in perfecting the precision and accuracy of measurements and developing new measurement tools and techniques, but in enhancing the conceptual richness of the field, and integrating theory-building and testing efforts across diverse science studies fields" (p. 3198).

Citation accuracy refers to the precise attribution of knowledge flow from cited to citing papers. Citation noise, on the other hand, introduces random variability into citation decisions, leading to inconsistencies and potential misrepresentation of research impact. In contrast to random variability of citation noise, citation bias represents systematic deviations, where certain factors influence citation decisions beyond actual knowledge flow, such as the reputation of authors or journals. A social citation system in which professional judgments of researchers can be seen as inconsistent, biased, and arbitrary loses credibility for its use in research evaluation. To analyze citation accuracy, citation bias, and citation noise, this study employs a fictitious social citation system, a simplified model of citing papers and cited papers. The system reveals the prevalence of citation errors and their possible influence on citation-based metrics. The statistical analysis of the system demonstrates that citation noise may lead to substantial distortions in citation impact measurements, undermining the reliability of these metrics in research evaluation.

This paper also explores strategies to mitigate citation noise and bias. The first strategy is the aggregation of citation decisions, leveraging the 'wisdom of crowds' to reduce the influence of individual-level random errors. However, aggregation does not correct systematic bias and is therefore not a general solution to citation inaccuracy. A second strategy is to improve citation decision hygiene through guidelines, training, citation justification tables, reduction to direct citations, post-publication citation corrections, and the use of AI. These measures aim to enhance the accuracy of citation decisions, ensuring that citations 'truly' reflect knowledge flow and contribute to more valid research evaluation.

This study underscores the importance of addressing citation noise and bias to ensure the validity and reliability of citation-based metrics. By understanding the complexities of citation dynamics and implementing targeted strategies, the scientific community can foster a more accurate and fair system of research evaluation. While citation bias has been extensively studied in bibliometrics, citation noise has received little attention. We argue that citation noise is at least as problematic as citation bias, as it introduces random variability into citation decisions, undermining the reliability of citation-based metrics. Citation noise is variability in citation decisions that should be identical between different authors. Variability is part of the scientific enterprise and part of citation decisions. However, when citation decisions are used to evaluate science, these decisions should be accurate. If they contain a lot of noise, they cannot measure knowledge flow.

Over the past decade, reform initiatives such as the Leiden Manifesto, the Metric Tide report, DORA, CoARA, and More Than Our Rank have promoted more careful and responsible uses of bibliometric indicators in research evaluation. These initiatives emphasize the limitations of citation-based metrics and caution against simplistic or excessive reliance on them. However, they have paid less attention to the citation decisions that generate citation data. From the perspective developed here, improving citation practices is one possible way to strengthen the validity of citation analysis.

This study has several limitations. First, the statistical analysis is based on a fictitious social citation system, which simplifies the complexities of real-world social citation systems. While this approach is useful for illustrating key concepts, it may not fully capture the dynamics of citation decisions in actual scientific communities. Second, the paper only partly provides empirical evidence for the prevalence of citation noise in real-world citation data. This evidence can only be derived, for example, from citation context studies. Future research should address this gap by conducting (large-scale) empirical studies to quantify the extent of citation noise and its impact on citation-based metrics. The intention of this conceptual paper was to introduce the general accuracy and noise framework for citation decisions.

Third, the discussion of citation noise and bias in this paper rests on the assumption that there is such a thing as a "correct" citation. From the perspective of improving the validity of citation analysis, it would be ideal if a universal citation norm existed to which all research communities subscribed. In practice, however, different research fields or research groups may use different but equally justifiable citation practices. As Wakeling et al. (2025) point out, "variation in [citation] error rates … may very well reflect the different standards by which those researchers are judging what constitutes an error, rather than error rates in an objective sense" (p. 1406). A further complication is the question of when authors should stop citing a particular work because its knowledge contribution has become so deeply incorporated into the canon that citation seems no longer necessary. This phenomenon is known as 'obliteration by incorporation' (Merton, 1965). For example, Fisher (1925) laid the groundwork for many of the statistical methods used in science today, yet this work is rarely cited. Obliteration by incorporation can be understood as part of a broader phenomenon: authors draw from "deeper or broader knowledge reservoirs than what is reflected in the paper's references" (Schilling & Green, 2011, p. 1325). The need to cite prior work is grounded in the cumulative nature of science, yet this same cumulative nature also implies that it may not be possible or desirable for authors to cite literally every source of every knowledge element present in their work.

To address the limitations of this paper, several avenues for future research can be proposed:

First, empirical studies are needed to investigate the prevalence and impact of citation noise in real-world citation data. Such studies could analyze the variability in citation decisions (noise) across different fields. The extent of citation noise can be measured in citation noise audits. These are experiments in which several experts in a field make independent citation decisions for the same scholarly text and variability in the decisions can be measured. Empirical studies are also needed to assess the extent to which noise affects citation-based metrics. Here, it would be also interesting to quantify how much citation pattern noise is stable and how much is citation occasion noise. These two types of pattern noise require different interventions to reduce the overall noise in citation decisions. Future research could also address heuristics (rules of thumb) authors use to cite certain papers (Bornmann & Marewski, 2019). If these heuristics are known, best-practice heuristics can be developed for ensuring accurate citation decisions.

Second, future research could investigate the number of citation decisions not only in favor of, but also against cited papers to determine the reliability of citations. Quantifying the reliability of citations requires data on citation decisions that did not lead to citations. One potential source of data for such studies is Mendeley (see https://www.mendeley.com). Mendeley is a reference management software that provides data on the number of users (readers) that added a certain work to their personal library. Research has shown that Mendeley reader counts can be used as indicators for the future use and citation of publications (Mohammadi et al., 2016; Thelwall, 2018). We assume that the difference between a publication's Mendeley reader counts and its number of citing authors can be used to estimate the number of citation decisions that did not manifest in citations. For such a comparison of counts and numbers, a problem must be solved: All citing co-authors would be counted. It is not known how many of the co-authors have a cited paper in their Mendeley library.

An alternative strategy to investigate citation noise and accuracy may be to perform empirical studies in rather small fields with a manageable amount of literature. Koffi (2025) proposes the 'omission indicator' to identify papers that should have been cited in citing papers. The 'omission indicator' is "a binary variable that detects when an article fails to cite a relevant previously published paper. Indeed, having determined the similarity score for every article pair [in a certain field], one can isolate the most similar articles to a given paper based on textual distance. Within this 'most similar set,' if a subsequent paper overlooks a preceding one despite several similarities, the omission indicator flags it with a value equal to one (or zero otherwise)" (p. 2208). This indicator may be used to estimate the accuracy of citation decisions and the reliability of citations in the respective field.

Third, future research could evaluate the effectiveness of the strategies proposed in this paper for reducing citation noise and bias (Lawson et al., 2025). Although we have presented measures for citation decision hygiene in this study, and other authors have presented similar measures for other areas, little is known about the effectiveness of these measures: “a fair amount of that is speculative – that is, it hasn’t been tested through research. It’s backed up indirectly – we didn’t completely invent things out of our head – but there is a lot of work that needs to be done to establish those things” (Kahneman & Frueh, 2022, p. 25). Thus, studies should assess the impact of citation guidelines, training programs, and AI-based tools on citation accuracy. These evaluations should consider not only the quantitative effects on citation metrics, but also the qualitative changes in citation practices and the perceptions of researchers regarding the role of citations in research evaluation. In this study, we focused on several strategies for improving citation accuracy. Other strategies are possible as well and could be elaborated in future studies such as the focus of evaluative citation analyses on journals working with measures to improve citation accuracy, for example, by demanding citation justification tables from the authors.

Fourth, future research could address possible contradicting effects of the strategies proposed above against citation noise and bias. On the one hand, AI-based tools may be helpful to improve the citation decision hygiene of authors. On the other hand, the use of the tools may lead to citation inflation in future papers whereby papers cite more work that shouldn’t be cited at all.

Fifth, the application of citation justification tables to promote citation decision hygiene could be tested in real-world settings. There may be barriers to its implementation, e.g., the additional burden it would place on time-poor researchers. If a journal starts to mandate such a table, it may reduce submissions, and therefore journals may be disincentivized to introduce the tables. Therefore, these tables could be implemented in academic writing workflows as a test, and the willingness to produce them and their effectiveness in reducing citation noise and improving the validity of citation-based metrics could be assessed through empirical studies. We could imagine that citation justification tables might be used in future studies to measure – based on the explanations of the citing

authors – the content of knowledge flow and the impact strength exercised by the cited paper on the citing publication.

# 5 Conclusions

Kahneman et al. (2021) provide a valuable framework for understanding the variability in human decisions. The framework can be reasonably applied in contexts including recurrent decisions (as opposed to single decisions). This is the case with citation decisions where interchangeable authors in a certain field make citation decisions in papers that are standardized (in style, content, structure etc.) using a pool of previous papers that are available to authors for possible citing. The existence of noise in citation decisions has significant implications for the integrity and reliability of research evaluation. By understanding and addressing the sources of noise, researchers and institutions can work towards creating a more consistent, fair, and meaningful bibliometric evaluation system. This not only enhances the quality of research evaluation processes but also strengthens trust in the academic enterprise. When citation decisions appear arbitrary or biased, trust in the academic evaluation system can diminish. By applying the framework proposed by Kahneman et al. (2021) to the context of citation decisions, we can identify strategies to mitigate noise and improve the overall rigor of academic citation practices.

This paper highlights the importance of addressing citation noise in citation analysis and argues that improving citation accuracy is essential for the reliable and valid use of citation-based metrics in research evaluation. According to Lawson et al. (2025), "problematic citing decisions give rise to a research literature that does not accurately represent the field's knowledge accumulation and unjustly discounts the work of less powerful and privileged researchers" (p. 3). The practical relevance of citation noise for bibliometric research evaluation depends on the level of aggregation. Performing bibliometrics on a high level of aggregation especially requires the assumption that citation decisions are unbiased. For example, citation data can be used to compare the scientific output of research institutions if the average number of citations for each institution is unbiased. Whereas the citation counts of large aggregates of publications are mostly subject to bias, the citation counts of individual papers may be subject to both bias and citation noise. The lower the level of aggregation, the more destructive are the potential consequences of citation noise. The bibliometric evaluation of individual papers or authors may only yield valid results if citation noise is below a critical threshold. While this paper provides a theoretical foundation for understanding citation analyses and proposes strategies to reduce citation noise, further research is needed to translate these ideas into practice. By addressing the limitations of this study and exploring the controversial aspects of citation noise, future research can contribute to a more accurate and equitable system of research evaluation.

# References

Aksnes, D. W., Langfeldt, L., & Wouters, P. (2019). Citations, citation indicators, and research quality: An overview of basic concepts and theories. *Sage Open*, *9*(1). https://doi.org/10.1177/2158244019829575

Algaba, A., Mazijn, C., Holst, V., Tori, F., Wenmackers, S., & Ginis, V. (2024). *Large language models reflect human citation patterns with a heightened citation bias*. Retrieved May 31, 2024, from https://arxiv.org/abs/2405.15739

Aman, V., & Gläser, J. (2025). Investigating knowledge flows in scientific communities: The potential of bibliometric methods. *Minerva*, *63*(1), 155-182. https://doi.org/10.1007/s11024-024-09542-2

American Psychological Association. (2020). *Publication manual of the American Psychological Association* (7. ed.). American Psychological Association (APA).

Barnes, S. B., & Dolby, R. G. (1970). The scientific ethos: A deviant viewpoint. *European Journal of Sociology/Archives Européennes de Sociologie*, *11*(1), 3-25.

Binz, M., Alaniz, S., Roskies, A., Aczel, B., Bergstrom, C. T., Allen, C., Schad, D., Wulff, D., West, J. D., Zhang, Q., Shiffrin, R. M., Gershman, S. J., Popov, V., Bender, E. M., Marelli, M., Botvinick, M. M., Akata, Z., & Schulz, E. (2025). How should the advancement of large language models affect the practice of science? *Proceedings of the National Academy of Sciences*, *122*(5), e2401227121. https://doi.org/10.1073/pnas.2401227121

Bornmann, L., & Daniel, H.-D. (2008). What do citation counts measure? A review of studies on citing behavior. *Journal of Documentation*, *64*(1), 45-80. https://doi.org/10.1108/00220410810844150

Bornmann, L., & Marewski, J. N. (2019). Heuristics as conceptual lens for understanding and studying the usage of bibliometrics in research evaluation. *Scientometrics*, *120*(2), 419-459. https://doi.org/10.1007/s11192-019-03018-x

Bruton, S. V., Macchione, A. L., Brown, M., & Hosseini, M. (2025). Citation ethics: An exploratory survey of norms and behaviors. *Journal of Academic Ethics*, *23*, 329-346. https://doi.org/10.1007/s10805-024-09539-2

Cawkell, A. E. (1969). Citation noise. *Aslib Proceedings*, *21*(11), 467.

Chen, B., Murray, D., Liu, Y., & Barabási, A.-L. (2024). *The origin, consequence, and visibility of criticism in science*. Retrieved December 19, 2024, from https://arxiv.org/abs/2412.02809

Donner, P., Stahlschmidt, S., Haunschild, R., & Bornmann, L. (2025). Does citation context information enhance the validity of citation analysis for measuring research quality? An empirical comparison of peer assessments and enriched citations. *Quantitative Science Studies*, *6*, 967–987. https://doi.org/10.1162/qss.a.15

Dougherty, M. R., Illingworth, D. A., & Nguyen, R. (2024). A memory-theoretic account of citation propagation. *Royal Society Open Science*, *11*(5). https://doi.org/10.1098/rsos.231521

Enkhbayar, A. (2025). *Citational matters: Towards a performative theory of citations*. Faculty of Communication, Art and Technology, Simon Fraser University.

Fisher, R. A. (1925). *Statistical methods for research workers*. Oliver & Boyd.

Galton, F. (1907). Vox populi. *Nature*, *75*, 450-451. https://doi.org/10.1038/075450a0

Gavras, H. (2002). Inappropriate attribution: The “lazy author syndrome”. *American Journal of Hypertension*, *15*(9), 831. https://doi.org/10.1016/s0895-7061(02)02989-8

Gigerenzer, G., & Brighton, H. (2009). Homo heuristicus: Why biased minds make better inferences. *Topics in Cognitive Science*, *1*(1), 107-143. https://doi.org/10.1111/j.1756-8765.2008.01006.x

Gilbert, G. N. (1977). Referencing as persuasion. *Social Studies of Science*, *7*(1), 113-122.

Hammarfelt, B., Rushforth, A., & de Rijcke, S. (2020). Temporality in academic evaluation: 'Trajectoral thinking' in the assessment of biomedical researchers. *Valuation Studies*, *7*, 33. https://doi.org/10.3384/VS.2001-5992.2020.7.1.33

Hicks, D., Wouters, P., Waltman, L., de Rijcke, S., & Rafols, I. (2015). Bibliometrics: The Leiden Manifesto for research metrics. *Nature*, *520*(7548), 429-431. https://doi.org/10.1038/520429a

Hulek, K., & Teschke, O. (2023). How do mathematicians publish? Some trends. *European Mathematical Society Magazine*, *129*, 36–41. https://doi.org/10.4171/MAG/160

Jaffe, A. B., Trajtenberg, M., & Fogarty, M. S. (2000). Knowledge spillovers and patent citations: Evidence from a survey of inventors. *American Economic Review*, *90*(2), 215-218. https://doi.org/10.1257/aer.90.2.215

Jannot, A. S., Agoritsas, T., Gayet-Ageron, A., & Perneger, T. V. (2013). Citation bias favoring statistically significant studies was present in medical research. *Journal of Clinical Epidemiology*, *66*(3), 296-301. https://doi.org/10.1016/j.jclinepi.2012.09.015

Jergas, H., & Baethge, C. (2015). Quotation accuracy in medical journal articles: A systematic review and meta-analysis. *PeerJ*, *3*, e1364. https://doi.org/10.7717/peerj.1364

Kahneman, D., & Frueh, S. (2022). "Try to design an approach to making a judgment; don't just go into it trusting your intuition". *Issues in Science and Technology*, *38*(3), 23-26.

Kahneman, D., Sibony, O., & Sunstein, C. R. (2021). *Noise: A flaw in human judgment*. Little, Brown.

Khan, S., Banu S, A., Pawde, A., Kumar, R., Akash, S., Dhama, K., & Amarpal, A. (2023). ChatGPT and artificial hallucinations in stem cell research: Assessing the accuracy of generated references – a preliminary study. *Annals of Medicine and Surgery*, *85*, 5275-5278. https://doi.org/10.1097/MS9.0000000000001228

Kim, L., Adolph, C., West, J. D., & Stovel, K. (2020). The influence of changing marginals on measures of inequality in scholarly citations: Evidence of bias and a resampling correction. *Sociological Science*, *7*, 314-341. https://doi.org/10.15195/v7.a13

Klitzing, N., Hoekstra, R., & Strijbos, J.-W. (2019). Literature practices: Processes leading up to a citation. *Journal of Documentation*, *75*(1), 62–77. https://doi.org/doi:10.1108/JD-03-2018-0047

Koffi, M. (2025). Innovative ideas and gender (in)equality. *American Economic Review*, *115*(7), 2207–2236. https://doi.org/10.1257/aer.20211811

Kousha, K., & Thelwall, M. (2024). Factors associating with or predicting more cited or higher quality journal articles: An Annual Review of Information Science and Technology (ARIST) paper. *Journal of the Association for Information Science and Technology*, *75*(3), 215-244. https://doi.org/10.1002/asi.24810

Lawson, K. M., Murphy, B. A., Azpeitia, J., Lombard, E. J., & Pope, T. J. (2025). Citing decisions in psychology: A roadblock to cumulative and inclusive science. *Advances in Methods and Practices in Psychological Science*, *8*(3), 1-22. https://doi.org/10.1177/25152459251351287

Masic, I. (2013). The importance of proper citation of references in biomedical articles. *Acta Informatica Medica*, *21*(3), 148-155. https://doi.org/10.5455/aim.2013.21.148-155

Meho, L., & Yang, K. (2007). Fusion approach to citation-based quality assessment. In D. Torres-Salinas & H. F. Moed (Eds.), *Proceedings of ISSI 2007: 11th international conference of the International Society for Scientometrics and Informetrics* (pp. 568–581). Centre for Scientific Information and Documentation (CINDOC), CSIC, Madrid, and ECOOM, KU Leuven.

Merton, R. K. (1965). *On the shoulders of giants: The post-italianate edition*. University of Chicago Press.

Merton, R. K. (1973). *The sociology of science: Theoretical and empirical investigations*. University of Chicago Press.

Milojević, S. (2025). Science of science. *Scientometrics*, *130*, 3195–3211. https://doi.org/10.1007/s11192-025-05322-1

Mogull, S. A. (2017). Accuracy of cited "facts" in medical research articles: A review of study methodology and recalculation of quotation error rate. *PLOS ONE*, *12*(9), e0184727. https://doi.org/10.1371/journal.pone.0184727

Mohammadi, E., Thelwall, M., & Kousha, K. (2016). Can Mendeley bookmarks reflect readership? A survey of user motivations. *Journal of the Association for Information Science and Technology*, *67*(5), 1198-1209. https://doi.org/10.1002/asi.23477

Murphy, E. J. (2011). Citations: The rules they didn't teach you. *Lipids*, *46*(4), 307-309. https://doi.org/10.1007/s11745-011-3543-3

Mutz, R., Wolbring, T., & Daniel, H.-D. (2017). The effect of the "very important paper" (VIP) designation in *Angewandte Chemie International Edition* on citation impact: A propensity score matching analysis. *Journal of the Association for Information Science and Technology*, *68*(9), 2139-2153. https://doi.org/10.1002/asi.23701

Nicolaisen, J. (2004). *Social behavior and scientific practice-missing pieces of the citation puzzle*. Royal School of Library and Information Science,

Nielsen, M. W., & Andersen, J. P. (2021). Global citation inequality is on the rise. *Proceedings of the National Academy of Sciences of the United States of America*, *118*(7). https://doi.org/10.1073/pnas.2012208118

Penders, B. (2018). Ten simple rules for responsible referencing. *PLOS Computational Biology*, *14*(4), e1006036. https://doi.org/10.1371/journal.pcbi.1006036

Pendlebury, D. A. (2025). *The legacy of Eugene Garfield: A pioneer of information science*. Retrieved October 28, 2025, from https://clarivate.com/academia-government/blog/the-legacy-of-eugene-garfield-a-pioneer-of-information-science/

Price, D. d. S. (1976). General theory of bibliometric and other cumulative advantage processes. *Journal of the American Society for Information Science*, *27*(5-6), 292-306. https://doi.org/10.1002/asi.4630270505

Rushforth, A., & Hammarfelt, B. (2023). The rise of responsible metrics as a professional reform movement: A collective action frames account. *Quantitative Science Studies*, *4*(4), 879-897. https://doi.org/10.1162/qss_a_00280

Schilling, M. A., & Green, E. (2011). Recombinant search and breakthrough idea generation: An analysis of high impact papers in the social sciences. *Research Policy*, *40*(10), 1321-1331. https://doi.org/10.1016/j.respol.2011.06.009

Small, H. (2004). On the shoulders of Robert Merton: Towards a normative theory of citation. *Scientometrics*, *60*(1), 71-79. https://doi.org/10.1023/B:SCIE.0000027310.68393.bc

Small, H. G. (1978). Cited documents as concept symbols. *Social Studies of Science*, *8*(3), 327-340.

Smith, D. (2014). Finding the signal in the noise of patent citations: How to focus on relevance for strategic advantage. *Technology Innovation Management Review*, *4*(9), 36-44. https://doi.org/10.22215/timreview/830

Sula, C. A., & Miller, M. (2014). Citations, contexts, and humanistic discourse: Toward automatic extraction and classification. *Literary and Linguistic Computing*, *29*(3), 452-464. https://doi.org/10.1093/llc/fqu019

Surowiecki, J. (2004). *The wisdom of crowds*. Random House.

Tahamtan, I., & Bornmann, L. (2018). Core elements in the process of citing publications: Conceptual overview of the literature. *Journal of Informetrics*, *12*(1), 203-216. https://doi.org/10.1016/j.joi.2018.01.002

Tahamtan, I., & Bornmann, L. (2019). What do citation counts measure? An updated review of studies on citations in scientific documents published between 2006 and 2018. *Scientometrics*, *121*(3), 1635–1684. https://doi.org/10.1007/s11192-019-03243-4

Tahamtan, I., & Bornmann, L. (2022). The Social Systems Citation Theory (SSCT): A proposal to use the social systems theory for conceptualizing publications and their citation links. *Profesional de la Información*, *31*(4), e310411. https://doi.org/10.3145/epi.2022.jul.11

Tang, B. L. (2023). Some insights into the factors influencing continuous citation of retracted scientific papers. *Publications*, *11*(4). https://doi.org/10.3390/publications11040047

Teplitskiy, M., Duede, E., Menietti, M., & Lakhani, K. R. (2022). How status of research papers affects the way they are read and cited. *Research Policy*, *51*(4), 104484. https://doi.org/10.1016/j.respol.2022.104484

Thelwall, M. (2018). Early Mendeley readers correlate with later citation counts. *Scientometrics*, *115*(3), 1231-1240. https://doi.org/10.1007/s11192-018-2715-9

Thelwall, M. (2024). *Quantitative methods in research evaluation: Citation indicators, altmetrics, and artificial intelligence*. Retrieved July 5, 2024, from https://arxiv.org/abs/2407.00135

Thelwall, M., Kousha, K., Stuart, E., Makita, M., Abdoli, M., Wilson, P., & Levitt, J. (2023). In which fields are citations indicators of research quality? *Journal of the Association for Information Science and Technology*, *74*(8), 941-953. https://doi.org/10.1002/asi.24767

Traag, V. A., & Waltman, L. (2022). *Causal foundations of bias, disparity and fairness*. Retrieved August 4, 2022, from https://arxiv.org/abs/2207.13665

van Raan, A. F. J. (2005). Fatal attraction: Conceptual and methodological problems in the ranking of universities by bibliometric methods. *Scientometrics*, *62*(1), 133-143. https://doi.org/10.1007/s11192-005-0008-6

Varga, A. (2022). The narrowing of literature use and the restricted mobility of papers in the sciences. *Proceedings of the National Academy of Sciences of the United States of America*, *119*(17). https://doi.org/10.1073/pnas.2117488119

Wakeling, S., Paramita, M. L., & Pinfield, S. (2025). How do authors perceive the way their work is cited? Findings from a large-scale survey on quotation accuracy. *Journal of the Association for Information Science and Technology*, *76*(10), 1396-1410. https://doi.org/10.1002/asi.70000

Wang, D., & Barabási, A.-L. (2021). *The science of science*. Cambridge University Press. https://doi.org/10.1017/9781108610834

Wang, H., Fu, T., Du, Y., Gao, W., Huang, K., Liu, Z., Chandak, P., Liu, S., Van Katwyk, P., Deac, A., Anandkumar, A., Bergen, K., Gomes, C. P., Ho, S., Kohli, P., Lasenby, J., Leskovec, J., Liu, T.-Y., Manrai, A., . . . Zitnik, M. (2023). Scientific discovery in the age of artificial intelligence. *Nature*, *620*(7972), 47-60. https://doi.org/10.1038/s41586-023-06221-2

Wei, F. F., Zhang, G. J., Zhang, L., Liang, Y. K., & Wu, J. B. (2019). Decreasing the noise of scientific citations in patents to measure knowledge flow. In G. Catalano, C. Daraio, M. Gregori, H. F. Moed, & G. Ruocco (Eds.), *17th international conference of the International Society for Scientometrics and Informetrics (ISSI) on Scientometrics and Informetrics* (pp. 1662-1669).

Wilsdon, J., Allen, L., Belfiore, E., Campbell, P., Curry, S., Hill, S., Jones, R., Kain, R., Kerridge, S., Thelwall, M., Tinkler, J., Viney, I., Wouters, P., Hill, J., & Johnson, B. (2015). *The metric tide: Report of the independent review of the role of metrics in research assessment and management*. Higher Education Funding Council for England (HEFCE).

Wouters, P. (1998). The signs of science. *Scientometrics*, *41*(1-2), 225-241.

Wouters, P. (1999). Beyond the holy grail: From citation theory to indicator theories. *Scientometrics*, *44*(3), 561-580.

# Appendix

Table 2. Fictitious social citation system with 1) no citation bias, 2) minimized citation system noise, 3) inaccurate citation counts for all cited papers ($TC_k \neq EC_k$).

| | Cited paper A | | | Cited paper B | | | Cited paper C | | | Cited paper D | | | | | | | |
|---|---|---|---|---|---|---|---|---|---|---|---|---|---|---|---|---|---|
| | R | A | E | R | A | E | R | A | E | R | A | E | $PR_{ij}$ | $PA_{ij}$ | $PE_{ij}$ | $\overline{PE_i}$ | $\sigma_{PN,i}$ |
| Author I: Citing paper 1 | 1 | 1 | 0 | 1 | 1 | 0 | 0 | 1 | 1 | 1 | 0 | 1 | 0.75 | 0.50 | 0.50 | | |
| Author I: Citing paper 2 | 1 | 1 | 0 | 1 | 1 | 0 | 0 | 1 | 1 | 1 | 0 | 1 | 0.75 | 0.50 | 0.50 | 0.50 | 0.00 |
| Author I: Citing paper 3 | 1 | 1 | 0 | 1 | 1 | 0 | 0 | 1 | 1 | 1 | 0 | 1 | 0.75 | 0.50 | 0.50 | | |
| Author II: Citing paper 4 | 1 | 1 | 0 | 1 | 1 | 0 | 0 | 1 | 1 | 1 | 0 | 1 | 0.75 | 0.50 | 0.50 | 0.50 | 0.00 |
| Author II: Citing paper 5 | 1 | 1 | 0 | 1 | 1 | 0 | 0 | 1 | 1 | 1 | 0 | 1 | 0.75 | 0.50 | 0.50 | | |
| Author III: Citing paper 6 | 1 | 0 | 1 | 0 | 1 | 1 | 1 | 1 | 0 | 1 | 1 | 0 | 0.75 | 0.50 | 0.50 | | |
| Author III: Citing paper 7 | 1 | 0 | 1 | 0 | 1 | 1 | 1 | 1 | 0 | 1 | 1 | 0 | 0.75 | 0.50 | 0.50 | | |
| Author III: Citing paper 8 | 1 | 0 | 1 | 0 | 1 | 1 | 1 | 1 | 0 | 1 | 1 | 0 | 0.75 | 0.50 | 0.50 | 0.50 | 0.00 |
| Author III: Citing paper 9 | 1 | 0 | 1 | 0 | 1 | 1 | 1 | 1 | 0 | 1 | 1 | 0 | 0.75 | 0.50 | 0.50 | | |
| Author III: Citing paper 10 | 1 | 0 | 1 | 0 | 1 | 1 | 1 | 1 | 0 | 1 | 1 | 0 | 0.75 | 0.50 | 0.50 | | |
| $PR_k$ | 1 | | | 0.50 | | | 0.50 | | | 1.00 | | | | | | | |
| $TC_k$ | 10 | | | 5 | | | 5 | | | 10 | | | | | | | |
| $EC_k$ | | 5 | | | 10 | | | 10 | | | 5 | | | | | | |
| $PA_k$ | | | 0.50 | | | 0.50 | | | 0.50 | | | 0.50 | | | | | |
| $PE_k$ | | | 0.50 | | | 0.50 | | | 0.50 | | | 0.50 | | | | | |
| $\overline{PA}$ | | | | | | | | | | | | | | 0.60 | | | |
| $\overline{PE}$ | | | | | | | | | | | | | | | 0.40 | | |
| $\sigma_{LN}$ | | | | | | | | | | | | | | | | 0.50 | |
| $\sigma_{PN}$ | | | | | | | | | | | | | | | | | 0.00 |

Notes. $R$ = realized citation, $A$ = accurate citation, $E$= erroneous citation decision, $PR_{ij}$ = proportion of realized citations of citing author $i$ in citing paper $j$, $PA_{ij}$ = proportion of accurate citations of citing author $i$ in citing paper $j$, $PE_{ij}$ = proportion of erroneous citations of citing author $i$ in citing paper $j$, $\overline{PE_i}$ = Person-specific average error rate of citing author $i$, $\sigma_{PN,i}$ = author-specific pattern noise (within-author variance of erroneous citations for author $i$), $PR_k$ = cited paper $k$'s proportion of realized citations, $TC_k$ = times cited, $EC_k$ = expected citations, $PA_k$ = cited paper $k$'s proportion of accurate citations, $PE_k$ = cited paper $k$'s proportion of erroneous citations, $\overline{PA}$ = Proportion of accurate citations in the entire citation system, $\overline{PE}$ = Proportion of erroneous citations in the entire citation system, $\sigma_{LN}$ = overall level noise (between-author variance of erroneous citations), $\sigma_{PN}$ = overall pattern noise.

Table 3. Fictitious social citation system with 1) no citation bias, 2) maximized citation system noise, 3) accurate citation counts for all cited papers ($TC_k = EC_k$).

| | Cited paper A | | | Cited paper B | | | Cited paper C | | | Cited paper D | | | Cited paper E | | | | | | | |
|---|---|---|---|---|---|---|---|---|---|---|---|---|---|---|---|---|---|---|---|---|
| | R | A | E | R | A | E | R | A | E | R | A | E | R | A | E | $PR_{ij}$ | $PA_{ij}$ | $PE_{ij}$ | $\overline{PE}_i$ | $\sigma_{PN,i}$ |
| Author I: Citing paper 1 | 1 | 0 | 1 | 1 | 0 | 1 | 1 | 0 | 1 | 1 | 0 | 1 | 1 | 0 | 1 | 1.00 | 0.00 | 1.00 | | |
| Author I: Citing paper 2 | 0 | 1 | 1 | 0 | 1 | 1 | 0 | 1 | 1 | 0 | 1 | 1 | 0 | 1 | 1 | 0.00 | 0.00 | 1.00 | 1.00 | 0.00 |
| Author I: Citing paper 3 | 1 | 0 | 1 | 1 | 0 | 1 | 1 | 0 | 1 | 1 | 0 | 1 | 1 | 0 | 1 | 1.00 | 0.00 | 1.00 | | |
| Author II: Citing paper 4 | 0 | 1 | 1 | 0 | 1 | 1 | 0 | 1 | 1 | 0 | 1 | 1 | 0 | 1 | 1 | 0.00 | 0.00 | 1.00 | | |
| Author II: Citing paper 5 | 1 | 0 | 1 | 1 | 0 | 1 | 1 | 0 | 1 | 1 | 0 | 1 | 1 | 0 | 1 | 1.00 | 0.00 | 1.00 | 1.00 | 0.00 |
| Author II: Citing Paper 6 | 0 | 1 | 1 | 0 | 1 | 1 | 0 | 1 | 1 | 0 | 1 | 1 | 0 | 1 | 1 | 0.00 | 0.00 | 1.00 | | |
| Author III: Citing paper 7 | 0 | 0 | 0 | 1 | 1 | 0 | 0 | 0 | 0 | 1 | 1 | 0 | 1 | 1 | 0 | 0.60 | 1.00 | 0.00 | | |
| Author III: Citing paper 8 | 0 | 0 | 0 | 1 | 1 | 0 | 0 | 0 | 0 | 1 | 1 | 0 | 1 | 1 | 0 | 0.60 | 1.00 | 0.00 | | |
| Author III: Citing paper 9 | 0 | 0 | 0 | 1 | 1 | 0 | 0 | 0 | 0 | 1 | 1 | 0 | 1 | 1 | 0 | 0.60 | 1.00 | 0.00 | 0.00 | 0.00 |
| Author III: Citing paper 10 | 0 | 0 | 0 | 1 | 1 | 0 | 0 | 0 | 0 | 1 | 1 | 0 | 1 | 1 | 0 | 0.60 | 1.00 | 0.00 | | |
| Author III: Citing paper 11 | 0 | 0 | 0 | 1 | 1 | 0 | 0 | 0 | 0 | 1 | 1 | 0 | 1 | 1 | 0 | 0.60 | 1.00 | 0.00 | | |
| $PR_k$ | 0.27 | | | 0.73 | | | 0.27 | | | 0.73 | | | 0.73 | | | | | | | |
| $TC_k$ | 3 | | | 8 | | | 3 | | | 8 | | | 8 | | | | | | | |
| $EC_k$ | | 3 | | | 8 | | | 3 | | | 8 | | | 8 | | | | | | |
| $PA_k$ | | | 0.45 | | | 0.45 | | | 0.45 | | | 0.45 | | | 0.45 | | | | | |
| $PE_k$ | | | 0.55 | | | 0.55 | | | 0.55 | | | 0.55 | | | 0.55 | | | | | |
| $\overline{PA}$ | | | | | | | | | | | | | | | | | 0.45 | | | |
| $\overline{PE}$ | | | | | | | | | | | | | | | | | | 0.55 | | |
| $\sigma_{LN}$ | | | | | | | | | | | | | | | | | | | 0.50 | |
| $\sigma_{PN}$ | | | | | | | | | | | | | | | | | | | | 0.00 |

Notes. $R$ = realized citation, $A$ = accurate citation, $E$= erroneous citation decision, $PR_{ij}$ = proportion of realized citations of citing author $i$ in citing paper $j$, $PA_{ij}$ = proportion of accurate citations of citing author $i$ in citing paper $j$, $PE_{ij}$ = proportion of erroneous citations of citing author $i$ in citing paper $j$, $\overline{PE}_i$ = Person-specific average error rate of citing author $i$, $\sigma_{PN,i}$ = author-specific pattern noise (within-author variance of erroneous citations for author $i$), $PR_k$ = cited paper $k$'s proportion of realized citations, $TC_k$ = times cited, $EC_k$ = expected citations, $PA_k$ = cited paper $k$'s proportion of accurate citations, $PE_k$ = cited paper $k$'s proportion of erroneous citations, $\overline{PA}$ = Proportion of accurate citations in the entire citation system, $\overline{PE}$ = Proportion of erroneous citations in the entire citation system, $\sigma_{LN}$ = overall level noise (between-author variance of erroneous citations), $\sigma_{PN}$ = overall pattern noise.

Table 4. Citation justification table (in chronological order from the beginning until the end of the manuscript)

| Cited work | Citation justification (knowledge used) |
|---|---|
| *Section: Introduction* | |
| Aman and Gläser (2025) | Scientific knowledge is a collective endeavor. |
| Enkhbayar (2025) | Definition of the term 'citation'. |
| Tahamtan and Bornmann (2022) | Science can be conceptualized as a social citation system. |
| Penders (2018) | Citations are a form of scientific currency, actively conferring or denying value. |
| Lawson et al. (2025) | Citations are, more than ever before, a currency of the scientific realm: They are a target. |
| Aksnes et al. (2019) | Citations can be interpreted as knowledge flow from cited to citing paper. |
| Merton (1973) | Rober K. Merton formulated a first theory of citation: the normative theory of citations. |
| Small (2004) | A norm of accurate acknowledgment of prior work may be expected in scholarly communication. |
| Penders (2018) | Well-referenced manuscripts place an argument in its proper knowledge context. |
| Lawson et al. (2025) | Citation should represent at least one specific input to the citing author's evidentiary and/or argument chain. |
| Bornmann and Daniel (2008) | Overview of the literature on reasons to cite and citation functions. |
| Tahamtan and Bornmann (2018) | Overview of the literature on reasons to cite and citation functions. |
| Tahamtan and Bornmann (2019) | Overview of the literature on reasons to cite and citation functions. |
| Gilbert (1977) | Example of reasons to cite. One of the most early identified reasons to cite in bibliometrics: Authors cite to persuade their readers. |
| Bornmann and Daniel (2008) | Further explanation of the reason why authors cite to persuade their readers. |
| Sula and Miller (2014) | Linguistics tends to feature reinforcing citations of prior literature, whereas philosophy typically involves more critical engagement with cited works. |
| Wakeling et al. (2025) | Overview of studies dealing with inaccurate citations. |
| Wakeling et al. (2025) | Definition of quotation errors. |
| Wakeling et al. (2025) | Definition of reference errors. |
| Gavras (2002) | Definition of the 'lazy author syndrome'. |
| Chen et al. (2024) | Definition of the 'heuristic citation approach'. |
| Barnes and Dolby | One of the first attempts towards a social-constructivist theory of |

| (1970) | citations. |
|---|---|
| Small (1978) | Conceptualization of citations as markers or symbols. |
| Wouters (1998, 1999) | Proposal of the reflexive indicator theory. |
| Nicolaisen (2004) | Conceptualization of references as threat signals. |
| Tahamtan and Bornmann (2022) | Proposal of the Social Systems Citation Theory (SSCT). |
| Tahamtan and Bornmann (2022) | A detailed overview of citation theories. |
| Kahneman et al. (2021) | Conceptualization of citation noise, citation bias, and citation accuracy which have been transferred to citation decisions. |
| Kahneman et al. (2021) | Conceptualization of citation noise, citation bias, and citation accuracy which have been transferred to citation decisions. |
| Kahneman and Frueh (2022) | Noise and bias also exist outside the justice system. |
| Traag and Waltman (2022) | Definition of bias. |
| Kousha and Thelwall (2024) | Overview of factors that can lead to biased citation decisions. |
| Jannot et al. (2013) | Example of citation bias: favoring statistically significant results in medicine. |
| Kahneman et al. (2021) | Conceptualization of citation noise, citation bias, and citation accuracy which have been transferred to citation decisions. |
| Cawkell (1969) | The term ‘citation noise’ appears in the title. |
| Meho and Yang (2007) | The term ‘citation noise’ appears in the abstract. |
| Tang (2023) | The term ‘citation noise’ appears in the abstract. |
| Wei et al. (2019) | The term ‘citation noise’ appears in the abstract. |
| Smith (2014) | Overview of the literature on citation noise in patent data. |
| Smith (2014) | Definition of knowledge flow. |
| Jaffe et al. (2000) | Citations in a patent not representing knowledge flow may account for half of its total citations. |
| Bornmann and Daniel (2008) | Studies on reasons to cite reveal a comparatively frequent occurrence of citations of the perfunctory (up to 50 percent) and persuasive (up to 40 percent) type. |
| Donner et al. (2025) | Around half of the citations in the WoS are of the background type. |
| Teplitskiy et al. (2022) | Most citations have little-to-no influence on the citing authors. |
| Wakeling et al. (2025) | Range of inaccurate citations reported in studies is from 5% to 40%. |
| Wakeling et al. (2025) | Most studies report inaccurate citations between 10% to 20%. |
| Wakeling et al. (2025) | Illustration of the results on inaccurate citations. |
| Jergas and Baethge (2015) | Meta-analysis of studies dealing with inaccurate citations reports that 25.4% of citations are inaccurate. |

| Mogull (2017) | Meta-analysis of studies dealing with inaccurate citations reports that 14.5% of citations are inaccurate. |
|---|---|
| Jaffe et al. (2000) | Citations in a patent do not represent knowledge flow in many cases. |
| Smith (2014) | Patent citation noise is seen as a major challenge to the effectiveness of patent evaluation methodologies, which may therefore be poor indicators of the economic value of patents. |
| Hulek and Teschke (2023) | Bibliometric data plays an important role in (almost) all evaluations. |
| Hammarfelt et al. (2020) | Bibliometrics plays an important role in the peer review process. |
| *Section: Definition and measurement of citation accuracy, citation bias, and citation noise* | |
| Gigerenzer and Brighton (2009) | The authors speak of accidental patterns. |
| Chen et al. (2024) | Examples of random citation decisions. |
| Kahneman et al. (2021) | The authors do not provide a mathematical formula for measuring citation occasion noise. |
| Kahneman et al. (2021) | Three types of noise are explained. |
| Jannot et al. (2013) | Citation bias favoring statistically significant results in medical research. |
| Mutz et al. (2017) | Papers labeled as very important papers by a reputable journal received more citations than comparable papers (i.e. papers of similar quality) without this label. |
| Kousha and Thelwall (2024) | Provide an overview of studies that investigated biases in citation decisions. |
| *Section: Strategies to reduce citation noise and citation bias* | |
| Rushforth and Hammarfelt (2023) | Introduction of the term “professional reform movement” for the reports and initiatives pushing for the careful and responsible use of bibliometric indicators. |
| Hicks et al. (2015) | Leiden Manifesto as an example of a professional reform movement. |
| Wilsdon et al. (2015) | Metric Tide report as an example of a professional reform movement. |
| Thelwall (2024) | Overview of professional reform movements. |
| Thelwall et al. (2023) | Reform movement delivers strong arguments against using citations for aspects of research assessment. |
| Rushforth and Hammarfelt (2023) | Responsible metrics movement has drawn attention to the risks of using citation metrics in research assessment. |
| Lawson et al. (2025) | Respondents agreed that ‘citing is foundational to the credibility of psychological scholarship’. |
| Penders (2018) | To cite well, or to reference responsibly, is a matter of concern to all researchers. |
| Smith (2014) | Long direct quote describing the emergence of citation noise in patents. |
| Smith (2014) | Citation noise in patent data may lead to insufficient indicators of the economic value of patents and may obscure thus valuable |

| | economic insights. |
|---|---|
| Bornmann and Daniel (2008) | Overview of empirical studies shows that half of the citations are of the perfunctory and persuasive type. |
| Wakeling et al. (2025) | Overview of empirical studies shows that 10% to 20% of citations are inaccurate. |
| Smith (2014) | Importance of accurate citations in patent analysis. |
| Murphy (2011) | Many authors are not selective about referencing papers and tend to be a bit overzealous in citing more and more papers, many of which are not on the topic. |
| Masic (2013) | Definition of plagiarism. |
| *Section: Aggregation of citation decisions* | |
| Galton (1907) | Introduction of the wisdom of crowds' principle. |
| Galton (1907) | Introduction of the wisdom of crowds' principle. |
| Pendlebury (2025) | How does Eugene Garfield describe his Science Citation Index? |
| Surowiecki (2004) | 'Wisdom of crowds' requires independent decisions |
| Price (1976) | The non-independence of citations has already been discussed in classical bibliometric literature. |
| Wang and Barabási (2021) | Explanation of preferential attachment/cumulative advantage |
| (Dougherty et al., 2024) | Cumulative advantage may explain citation concentration |
| Kim et al. (2020) | Cumulative advantage may explain citation concentration |
| Nielsen and Andersen (2021) | Cumulative advantage may explain citation concentration |
| Varga (2022) | Cumulative advantage may explain citation concentration |
| van Raan (2005) | Method of aggregation targets citation errors. |
| Kahneman et al. (2021) | Aggregated decisions may be less noisy, but not less biased than the individual judgments. |
| *Section: Citation decision hygiene* | |
| Lawson et al. (2025) | Empirical results from a survey. |
| *Section: Guidelines for ensuring accurate citations* | |
| Wakeling et al. (2025) | Many citations do not correspond to the content of a previous work or do not support the statement it accompanies. |
| Lawson et al. (2025) | Citing behaviors can increasingly be understood through the lens of satisficing. |
| Wakeling et al. (2025) | Reviewers and editors should be instrumental in detecting citation errors. |
| Klitzing et al. (2019) | Citation guidelines do not provide any guidance on when it is necessary to cite a particular publication and when it is not. |
| American Psychological Association (2020) | Guideline for accurate citing. |
| Lawson et al. (2025) | Critique of the citation guidelines of the American Psychological Association. |
| Kahneman and Frueh (2022) | Structuring processes is the best way to improve human decision-making. |
| Bruton et al. (2025) | Results of a survey on ethically questionable citation behaviors. |
| Bruton et al. (2025) | Study shows that ethical norms of citations are rarely articulated clearly and meticulously. |
| Murphy (2011) | One of the few guidelines about citing accurately. |
| Murphy (2011) | List of eight rules in the citation guidelines refers to different aspects in connection with citation decisions but essentially concerns the indication of knowledge flow. |
| Penders (2018) | Citations should be used to differentiate own research from that of others, and citations point readers to relevant literature. |
| Penders (2018) | Citations should be used to differentiate own research from that of others, and citations point readers to relevant literature. |
| Penders (2018) | Use of superficial citations is opposed. |
| *Section: Citation justification table* | |
| Lawson et al. (2025) | Journals should request from authors an annotated reference section with approximately one to two sentences for why each cited work was included in the manuscript. |
| Penders (2018) | Citations in publications should be explained. |
| *Section: Training of (early-career) researchers* | |
| Lawson et al. (2025) | Outlining the problem of the lack of training in psychology. |
| Lawson et al. (2025) | Results from the survey of psychology-journal editorial board members. |
| American Psychological Association (2020) | Example of a publication manual. |
| *Section: Correction of citation errors* | |
| Wakeling et al. (2025) | What measures authors took when they had ever come across a citation of their work that they considered inappropriate. |
| Wakeling et al. (2025) | What measures authors took when they had ever come across a citation of their work that they considered inappropriate. |
| *Section: The use of AI for citation decision correction* | |
| Binz et al. (2025) | Overview of how AI could be integrated into the work process of researchers. |
| Wang et al. (2023) | Overview of how AI could be integrated into the work process of researchers. |
| Algaba et al. (2024) | Study investigating the use of AI to suggest scholarly references. |
| Algaba et al. (2024) | LLMs can aid in citation generation but may also generate biases. |
| Khan et al. (2023) | ChatGPT can produce generally accurate references but occasionally generates artificial hallucinations. |
| Khan et al. (2023) | Study investigating the use of AI to suggest scholarly references. |
| Algaba et al. (2024) | Study investigating the use of AI to suggest scholarly references. |
| *Section: Discussion* | |
| Aman and Gläser (2025) | Research papers include codified knowledge that is addressed to a large and partly anonymous audience. |
| Milojević (2025) | Call for perfecting the precision and accuracy of measurements and developing new measurement tools and techniques in diverse science studies fields. |
| Wakeling et al. (2025) | Different standards by which researchers are judging what constitutes citation errors. |

| Merton (1965) | Definition of ‘obliteration by incorporation’. |
|---|---|
| Fisher (1925) | The author introduced the statistical method ‘variance analysis’. |
| Schilling and Green (2011) | Authors draw from a deeper or broader knowledge reservoirs than what is reflected in a paper’s cited references. |
| Mohammadi et al. (2016) | Mendeley reader counts can be used as indicators for the future use of publications. |
| Thelwall (2018) | Mendeley reader counts can be used as indicators for the future citations of publications. |
| Koffi (2025) | Proposes the ‘omission indicator’ to identify papers that should have been cited in citing papers. |
| Lawson et al. (2025) | Future research could evaluate the effectiveness of the strategies for reducing citation errors. |
| Kahneman and Frueh (2022) | Future research should evaluate the effectiveness of strategies for reducing noise in human decision-making. |
| *Section: Conclusions* | |
| Kahneman et al. (2021) | Providing a framework for understanding the variability in human decisions. |
| Kahneman et al. (2021) | The framework proposed by Kahneman et al. (2021) has been applied to the context of citation decisions. |
| Lawson et al. (2025) | Problematic citing decisions give rise to a research literature that does not accurately represent the field’s knowledge accumulation. |

## Contributors’ roles

Conceptualization: LB, CL
Formal analysis: LB, CL
Investigation: LB, CL
Methodology: LB, CL
Project administration: LB
Software: LB, CL
Supervision: LB
Validation: LB, CL
Writing – original draft: LB, CL
Writing – review & editing: LB, CL

## Funding source statement

No funding was received for conducting this study.

## Competing interest statement

The authors have no competing interests to declare that are relevant to the content of our paper.

# Declaration of generative AI in scientific writing

During the preparation of this work the authors used Gemini and ChatGPT to improve language and readability. After using this tool, the authors reviewed and edited the content as needed and take full responsibility for the content of the publication.

# Acknowledgement

We thank Adrian Barnett, Paul Donner, Robin Haunschild, Rüdiger Mutz, and the two reviewers Simon Wakeling and Paul Wouters for their very helpful feedback on our manuscript. The assessments of the reviewers were conducted via the MetaROR platform: https://doi.org/10.70744/MetaROR.356.2.ea. LB is member of the Distinguished Reviewers Board of *Scientometrics*.